\documentclass[12pt,preprint]{aastex}
\def\kms{\rm{km \ s^{-1}}}
\def\etal{\rm{et al. }}
\def\deg{^\circ }

\tighten
\begin{document}

\title{VLA HI Observations of Gas Stripping 
in the Virgo Cluster Spiral NGC~4522}
\author {Jeffrey D. P. Kenney\altaffilmark{1}
\& J. H. van Gorkom\altaffilmark{2}
\& Bernd Vollmer \altaffilmark{3}}
\email{kenney@astro.yale.edu, jvangork@astro.columbia.edu, bvollmer@newb6.u-strasbg.fr}
\altaffiltext{1}{Yale University Astronomy Department, P.O. Box 208101, New Haven, CT 06520-8101 USA}
\altaffiltext{2}{Department of  Astronomy, 
Columbia University, 550 West 120th Street, New York, NY 10027}
\altaffiltext{3}{CDS, Observatoire Astronomique de Strasbourg, UMR 7550, 11 Rue de l'Universite, 67000
 Strasbourg, France}

\received{someday}

\shorttitle{Gas Stripping in Virgo Spiral NGC~4522}
\shortauthors{Kenney, van Gorkom \& Vollmer}
\begin{abstract} 
We present VLA HI observations at $\sim$20$''$$\simeq$1.5 kpc resolution
of the highly inclined, HI-deficient Virgo cluster spiral galaxy NGC~4522, which 
is one of the clearest and nearest cases of ongoing ICM-ISM stripping.
HI is abundant and spatially coincident with the stellar disk in the center,
but beyond R = 3 kpc the HI distribution in the disk is sharply truncated
and the only HI is extraplanar, and all on the northwest side.
Forty percent of the total HI, corresponding to 1.5$\times$10$^8$ M$_{\sun}$, is extraplanar
and has likely been removed from the galaxy disk by an ICM-ISM interaction.
The kinematics and the morphology of the HI 
appear more consistent with ongoing stripping,
and less consistent with gas fall-back which may occur long after peak pressure.
Some of the extraplanar gas has linewidths (FWZI) of 150 km s$^{-1}$, 
including a blueshifted tail of weaker emission, and
much of the extraplanar gas exhibits a modest net blueshift with respect to the 
galaxy's disk rotational velocities, 
consistent with gas accelerated toward the  mean cluster velocity.
The SW side of the galaxy has less HI in the disk but more HI in the halo,
suggesting more effective gas removal on the side of the galaxy 
which is rotating into the ICM wind.
In recent simulations of ICM-ISM interactions, 
large surface densities of extraplanar gas like that in NGC~4522
are seen at relatively early stages of active stripping,
and not during later gas fall-back stages.
The galaxy is 3.3$\deg$$\simeq$800 kpc from M87, somewhat outside the
region of strongest cluster X-ray emission.
The ram pressure at this location, 
assuming a static smooth ICM and standard values for ICM density
and galaxy velocity, appears inadequate to cause the observed stripping.
We consider the possibility that the ram pressure is significantly stronger 
than standard values, 
due to large bulk motions and local density enhancements of the ICM gas,
which may occur in a dynamic, shock-filled ICM experiencing subcluster merging.
The HI and H$\alpha$ distributions are similar,
with both truncated in the disk at the same radius and HII regions 
located throughout much of the extraplanar HI.
This implies that the star-forming molecular ISM has been effectively 
stripped from the outer disk of the galaxy along with the HI. 
The inferred peak stripping rate of $\sim$10 M$_{\sun}$ yr$^{-1}$
is much larger than the galaxy's total star formation rate of 
$\sim$0.1 M$_{\sun}$ yr$^{-1}$,
implying that the rate of triggered star formation due to ICM pressure
is presently minor compared to the rate of gas lost due to stripping.
\end{abstract}

\keywords{
galaxies: ISM --- 
galaxies: interactions  --- 
galaxies: clusters: general --- 
galaxies: clusters: individual (Virgo)  --- 
galaxies: evolution --- 
galaxies: peculiar --- 
galaxies: structure} 
\section {Introduction}

Interactions between the interstellar medium (ISM) of a galaxy and
the gas in the intracluster medium (ICM) are believed to be
one of the main processes which drive galaxy evolution in clusters
(Gunn \& Gott 1972; Poggianti \etal 1999; van Gorkom 2003; Koopmann \& Kenney 2004).
Gas stripping is likely responsible for the transformation of many cluster
spirals into  S0's and Sa's, and thus may help cause both the morphology-density
relation (Dressler 1980)
and the Butcher-Oemler effect (Butcher \& Oemler 1978; Dressler etal 1997).
ISM stripping is also significant for the evolution of the ICM, as
gas removed from the galaxies enriches the ICM with heavy elements.

There is widespread indirect evidence for  ICM-ISM stripping in clusters
from single dish HI observations which show many HI-deficient galaxies
(e.g.  Giovanelli \& Haynes 1983; Helou etal 1984; Solanes etal 2002) 
and resolved observations which show
truncated HI (Warmels 1988; Cayatte etal 1990) and H$\alpha$ gas disks 
(Koopmann \& Kenney 1998; 2004)
in spiral galaxies with fairly normal stellar disks.
For most of these galaxies, it is not
clear whether stripping is active or occurred in the past.
There are several examples of cluster galaxies
which are likely to be undergoing active ICM-ISM stripping,
including some in Virgo 
(Phookun \& Mundy 1995; Vollmer etal 2001b; 
Yoshida etal 2002; Tschoke etal 2001),
Coma (Bravo-Alfaro \etal 2000, 2001; Vollmer etal 2001a)
and the A1367 clusters (Gavazzi et al.~1995). 
However there are few nearby galaxies which are unambiguously experiencing
ram pressure stripping from an ICM-ISM interaction, and which can be studied
with good sensitivity and high resolution.

Perhaps the clearest and 
nearest case of a cluster spiral with an ongoing ICM-ISM interaction
is the highly inclined Virgo galaxy NGC~4522.
The R-band and H$\alpha$ imaging of Kenney \& Koopmann (1999)
revealed an undisturbed stellar disk and a selectively disturbed
ISM, with extraplanar filaments of ionized gas containing HII regions
arising from the edge of a truncated star forming disk.
Based on the optical data Kenney \& Koopmann conclude that NGC~4522 is experiencing
active stripping of its ISM by an interaction with the ICM.
In this paper we present new HI and radio continuum images of NGC~4522
from the Very Large Array (VLA), which seem to confirm ongoing
ICM-ISM stripping,
and allow some of the details of this process to be studied.

With detailed observations of a galaxy experiencing ICM-ISM stripping
we can begin to learn what actually happens in an ICM-ISM interaction,
the physics of the interaction,
and the consequences for galaxy evolution.
Is there a shock, can the shock front be identified,
and does the shock front partially protect the galaxy?
How does the complex multi-phase ISM respond to an interaction
with the ICM, and what simplifying assumptions are reasonable
for modelling?
How do molecular clouds, which may be too dense to strip directly,
respond to the ICM pressure?
What is the acceleration profile of the stripped gas?
Can gas be pushed outwards from the local disk, but still be bound
to the halo of the galaxy? Does gas pushed outwards sometimes
fall back onto the galaxy?
When does the HI become ionized, and by what mechanism?
Is star formation triggered by ICM pressure, and if so when and where 
and under what conditions? 
How do the rates of triggered star formation and gas removal
by stripping compare?

In this paper, we address some of these issues, using the detailed
morphological and kinematic information from the new VLA HI data.
The paper is organized as follows. In \S 2 we introduce the galaxy NGC~4522.
In \S 3 we discuss the observations.
In \S 4 
we present the HI data, discuss morphological and kinematic properties
of the HI which are likely related to the interaction, and compare the HI with
other ISM tracers.
In \S 5 we discuss the interpretation of the HI properties in terms of an
ICM-ISM interaction, and compare the observations with simulations and
theoretical expectations.
We also compare NGC~4522 with other galaxies which are likely experiencing ICM-ISM interactions.
A summary is presented in \S 6.
Radio continuum maps of NGC~4522 are presented in a companion paper by Vollmer etal (2004).

\section {The Galaxy NGC~4522}

NGC~4522 is a highly inclined, medium-sized, spiral galaxy with a small
bulge-to-disk ratio. 
Assuming a Virgo cluster distance of 16 Mpc 
(Yasuda etal 1997; Kelson etal 2000; Solanes \etal 2002),
its apparent blue magnitude of 12.0 implies an absolute magnitude of -18.9,
corresponding to 0.5L$^*$.
Given a maximum rotation velocity of 103 km s$^{-1}$ (Rubin etal 1999),
the dynamical mass
is 2.0$\times$10$^{10}$ M$_{\sun}$ within the optical isophotal radius
of R$_{25}$=1.85$'$=8.6 kpc.
Strong dust lane extinction is observed to the NW of the major axis,
suggesting that the SW is the near side.

The optical peculiarities of NGC~4522 are reflected in its classification
of SBcd:(sp) (RC3) and Sc/Sb: (Binggeli, Sandage, \& Tammann 1985). 
The uncertain Sc/Sb Hubble classification probably results
from the small bulge and the relatively high star formation rate in the 
inner region (like Sc), in combination with the lack of star formation 
in the outer
disk (more like Sb). The galaxy is highly inclined (i=75$\pm$5$\deg$),
making it possible to see extraplanar features.
A photograph in Sandage \& Bedke (1994) shows an ``arm fragment''
emerging from the disk, which we showed in Kenney \& Koopmann (1999)
to be composed primarily of HII regions.
It is significant that all the known peculiarities in NGC~4522 are
associated with dust, gas, or HII regions, and not older stars.
The deep broadband optical images of Kenney \& Koopmann (1999)
show no evidence of any significant peculiarity in the stellar distribution.
The outer disk has
regular elliptical isophotes all at the same position angle as the inner galaxy.
These optical data clearly show an undisturbed stellar disk.

Single dish observations with the Arecibo telescope (Helou \etal 1984),
indicate an HI deficiency of 0.6,
meaning that NGC~4522 contains only $\sim$1/4 the amount
of atomic gas that a normal late type spiral of its size would contain
(Giovanelli \& Haynes 1983; Kenney \& Young 1989).
It was not included in either the 
Westerbork (Warmels 1988) or VLA (Cayatte \etal 1990) 
Virgo cluster HI imaging surveys.
The galaxy has a line-of-sight velocity of 2337 km s$^{-1}$, which is well
within the -400 to 2800 km s$^{-1}$ range occupied by Virgo cluster galaxies
(Binggeli, Popescu, \& Tammann 1993; Conselice etal 2001).
It is 3.3$\deg$ from M87, which is at the center of the main galaxy 
concentration in Virgo, 
and only 1.5$\deg$ from M49, which is at center of subcluster B
(Binggeli, Popescu, \& Tammann 1993).
Binggeli, Sandage, \& Tammann (1985) consider it a member of the Virgo
cluster,
and Yasuda \etal (1997) and Solanes etal (2002) find a Tully-Fisher distance 
consistent with Virgo cluster membership.
This part of the Virgo cluster contains many HI-deficient
galaxies (Haynes \& Giovanelli 1986; Solanes etal 2002),
as well as X-ray emission from hot intracluster gas (Bohringer \etal
1994).
A compilation of galaxy properties is given in Table 1 of Kenney \& Koopmann
(1999).





%

\section {Observations}

HI 21cm line observations of NGC~4522  were made on 20 March 2000,
using the CS array of the Very Large Array (VLA) of the National Radio Astronomy Observatory
\footnote{
The National Radio Astronomy Observatory is a facility of the National Science Foundation 
operated under cooperative agreement by Associated Universities, Inc.}
which is similar to the standard C array, 
but includes some shorter baselines (the spacings range from 0.035 to 3.4 km).
NGC~4522 was observed for 6.7 hours.
The nearby phase calibrators 1254+116 (0.74 Jy) or 1331+305 = 3C286
were observed every 30 minutes, and
1331+305 (14.8 Jy) was used as a flux and passband calibrator.
For calibration and mapping, we used the NRAO Astronomical Image Processing 
System (AIPS). 

The digital correlator was configured to produce
127 channels, two polarizations and used no hanning smoothing.
The resulting channel spacing is 5.2 km s$^{-1}$, and the
total bandpass is 655  km s$^{-1}$. 
The bandpass was centered at 2330  km s$^{-1}$, 
close to the heliocentric velocity of 2337  km s$^{-1}$
obtained by Rubin etal (1999).

To subtract the continuum we made a linear fit to the visibilities of 
the ``line free'' channels with the AIPS task UVLIN.
Channel maps were made  with the AIPS IMAGR task, using
CLEANing to remove sidelobes. 
The HI results presented in this paper are based on a data cube 
made with natural weighting, which optimizes surface brigthness
sensitivity, resulting in 
a beam of  22.5$''$$\times$16.4$''$ at PA =-43$\deg$. 
This produced the best combination of good spatial resolution and detected flux.
Adjacent velocity channels were averaged to yield an effective resolution
of 10.4 km s$^{-1}$. Twenty-five channels of this cube
contained emission greater than 3$\sigma$.
Figure 1 shows channel maps of the central 30 channels.
The rms in these channel maps is 0.45 mJy beam $^{-1}$,
and the peak signal-to-noise ratio is 32.

We detect a flux of 6.3$\pm$0.5 Jy km s$^{-1}$ which is only marginally less 
than the 7.0$\pm$1.0 Jy km s$^{-1}$ detected with Arecibo telescope
(Helou et al 1984). We made tapered images to search for any very
extended HI that the interferometer could have missed, but did not
detect any.

Continuum maps made from the 
``line-free'' channels are presented in Vollmer etal (2004).

\section {Results}

\subsection {HI Morphology} 

An image of the total HI intensity is shown in Figure 2.
This was obtained by summing the channel images which showed HI 
emission, after blanking pixels with values below 2$\sigma$
after spatial smoothing.
An HI intensity map superposed 
on an optical image of NGC~4522 in Figure 3 shows that
HI is abundant and spatially coincident with the stars
in the central 40$''$=3.1 kpc = 0.4R$_{25}$ of the galaxy.
In the plane of the stellar disk, the HI distribution is truncated
beyond R=0.4R$_{25}$. 
Beyond this radius, the only HI is located above the disk plane,
and all of it is on the NW side.
HI extends 60$''$=4.5 kpc above the disk plane.

In the following we refer to emission within -20$''$ to +11$''$
of the major axis as ``disk'' emission, and +11$''$ to +50$''$
above the major axis as ``extraplanar'' emission,
although this is an oversimplification since the region from
+11$''$ to +25$''$ contains both disk and extraplanar emission.
We have also made analyses with  a symmetric -11$''$ to +11$''$ region as a disk component,
and find very similar results. We show the  -20$''$ to +11$''$ region in
this paper since it includes more flux.
The ``disk'' and ``extraplanar'' regions are outlined in 
Figure 4, which shows the HI moment 0 map rotated by 57 degrees so that
the major axis is oriented horizontally.

The amount of extraplanar HI 
has been estimated by assuming that the disk emission is symmetric about the
major axis, and subtracting a disk contribution from 
regions to the northwest of the major axis.
The extraplanar HI comprises 40$\pm$8\% of the total HI emission, or
1.5$\times$10$^8$ M$_{\sun}$.
The ratio of extraplanar to disk gas differs on the 2 sides of the 
minor axis --in the SW it is 1.3 while in the NE it is 0.7.

Within the stellar disk, the HI distribution is asymmetric,
with 1.5 times more emission on the (lower-velocity) NE side than 
the (higher-velocity) SW side. 
This difference is illustrated in Figure 5, which shows
HI intensity slices along the disk and extraplanar components.

The extraplanar components also exhibit spatial asymmetries,
as shown in Figures 4 and 5.
The projected HI surface density distribution exhibits secondary peaks
25-30$''$=1.9-2.3 kpc above the disk plane. 
The highest surface brightness component is the SW component,
40$''$ from the minor axis.
Its projected HI surface density is 10 M$_{\sun}$ pc$^{-2}$,
a remarkably high value for extraplanar gas,
and not much less than the peak disk value of  16 M$_{\sun}$ pc$^{-2}$.
Toward the NE, the brightest extraplanar component is
70$''$  from the minor axis.
The surface brightness of the SW extraplanar peak is twice that of the 
NE peak, and the HI emission 
extends ~25\% further from the minor axis toward NE than toward the SW.
A third weaker extraplanar HI component 
lies a bit (0-15$''$) NE  of the minor axis,
extending 10-50$''$ above the disk.
Thus, the total extraplanar HI emission 
shows a modest excess in the SW (ratio NE/SW =0.8), but is stretched out more in the NE.

Figure 5 shows that
while the disk and extraplanar HI have very different radial distributions
on the 2 sides of the minor axis, 
the radial distributions and amounts of the total HI (disk plus extraplanar)
are remarkably similar on the 2 sides,  out to 60$''$.
Moreover, the total radial distribution profiles vary smoothly out to 60$''$,
showing no discontinuity in profile shape at r$\simeq$40-50$''$, where the disk
emission is truncated and the strongest extraplanar emission exists.
Both facts suggest that the extraplanar HI originated recently from nearby disk positions.

The HI deficiency of the galaxy is 0.6, suggesting that 
75\% of the original HI content has been lost from the galaxy,
and has likely been ionized. 
Of the hypothetical pre-stripped HI content,
10\% is extraplanar gas which has been removed from
the disk but still exists as HI, and
another 15\% remains as HI in the disk plane in the central 0.4R$_{25}$
of the galaxy.

\subsection {HI Kinematics} 

Figure 6 shows the HI spectral line profile for the entire galaxy,
as well as the disk and  extraplanar components.
The global profile shows a modest velocity asymmetry, with
emission at lower velocities 20\% stronger than those of
higher velocities.
This asymmetry is due to a strong NE/SW asymmetry in the disk component,
as seen in the total intensity distributions in Figure 4 and 5.
The extraplanar component is more symmetric,
reflecting the similar amounts of HI on the 2 sides of the minor axis.
While the velocity extents of the strongest disk and extraplanar components
are similar, the extraplanar component contains a blueshifted tail of emission at velocities 
$\sim$50 km s$^{-1}$ lower than any disk velocity.

An intensity weighted velocity field is shown in Figure 7.
This was made with a slightly higher flux cutoff (2.5$\sigma$),
than the total intensity map, only including emission that we trusted
to be real based on coherence of features either spatially or in velocity.
Figure 7 shows that right above the disk, both the SW and NE extraplanar components have
less extreme velocities than the nearby disk emission,
so that the isovelocity contours curve away from the minor axis as one goes from the
disk to the extraplanar gas.

Such a decrease can occur from projection effects
in galaxies which are not fully edge-on.
Also, in a few well studied cases of nearby non-cluster edge-on spirals,
evidence has been found for a
decrease in rotation velocity in gas above and below the disk
(Swaters etal 1997; Fraternali etal 2002).
However, what we see in NGC~4522 is different from what is seen in these galaxies.
The velocities change gradually in going from the disk to extraplanar emission over part of the
galaxy, but sharply at large radii in the SW.
Also in the SW the extraplanar galactocentric velocities first drop, then rise again, resulting in
isovelocity curves which resemble backwards question marks.

A map of the velocity dispersion is shown in Figure 8,
made using the same blanked cube that was used for the velocity field.
Most galaxies show peak linewidths near their galactic centers,
due to both larger intrinsic dispersions and the effects of beam-smearing where there are
large velocity gradients in the velocity field.
However, in NGC~4522, the peak is not at the center, but $\sim$30$''$ from the nucleus, in
the SW extraplanar gas, where the peak FWHM is 45  km s$^{-1}$.
Even within the disk, the maximum value is not at the nucleus but $\sim$7$''$ SW of the nucleus.

Position-velocity (PV) diagrams of the HI are shown in Figures 9, 10 and 12,
for cuts both parallel and perpendicular to the major axis.
The regions covered in the PV diagrams are indicated by the
inset boxes on the total intensity HI maps in Fig 4 and Fig 11.

The major axis PV plot of the disk component in Figure 9 shows
that despite the large spatial asymmetry of HI in the disk,
the disk HI kinematics are  relatively symmetric and normal.
The rotation curve indicated by the HI is similar to the H$\alpha$ 
rotation curves of Rubin etal (1999) and Vollmer etal (2000).
From symmetry considerations, the HI systemic velocity is determined
to be 2337$\pm$3 km s$^{-1}$, in excellent agreement with the optical
H$\alpha$ emission line value of
2337 km s$^{-1}$ from Rubin etal (1999).
The location of the kinematic center is 
RA = 12$^{\rm h}$33$^{\rm m}$39.64$^{\rm s}$ DEC = +09$\deg$10$'$27.2$''$
(J2000). This agrees well, within 1$''$, with the location of peak intensity
in the radio continuum map (Vollmer etal 2004).

A comparison of the disk and extraplanar
spatial velocity diagrams in Figure 10 shows several 
interesting features. 
1. Gas velocities extend to lower (blueshifted) values, but not to higher values
  in the extraplanar component. Whereas the brightest extraplanar component
  in the NE has a velocity outside the range of (lower in absolute sense) 
  any in the disk, the  brightest extraplanar component
  in the SW has a velocity similar to, and a bit lower than, 
   those found in the nearby disk.
2. There are larger linewidths in the extraplanar component, especially in the SW.
3. In the SW part of the galaxy, the extraplanar component
   has velocities which range from those found in the nearby disk, 
   to velocities $\sim$100 km s$^{-1}$ lower.
   In contrast, in the NE part of the galaxy from r=10-40$''$
   the extraplanar component has a modest redshift of 10-20 km s$^{-1}$ with 
   respect to the disk emission.

Some of the kinematic differences between the disk and extraplanar emission
are further illustrated in the spatial velocity diagrams in Figure 12, which are cuts perpendicular to
the major axis.  The large linewidth components in the SW extraplanar gas are seen
especially in Figures 12 f and g. Fig 12g shows the cut through the major SW extraplanar component,
as well as the nearby disk. One can clearly see the lower peak velocity of the extraplanar component,
relative to the disk emission, as well as its broad asymmetric line with FWZI=150 km s$^{-1}$.

\subsection {Comparion of HI with Other ISM tracers}


The neutral hydrogen distribution in NGC 4522 is quite remarkable. Although
we have a somewhat incomplete knowledge of the distribution of other components
of the ISM, current data are consistent with its entire ISM being
distributed roughly like the HI, with virtually no gas in the outer disk
and all ISM phases in the extraplanar component.

A comparison of HI with H$\alpha$ (Kenney \& Koopmann 1999)
is shown in Figure 13.
The spatial distributions of these two ISM components are similar
on scales $\geq$15$''\simeq$1 kpc,
although there are differences on smaller scales, similar to those seen in most galaxies.
HI and H$\alpha$ are co-extensive within the central 0.4R$_{25}$ 
of the optical disk. The H$\alpha$ intensity is asymmetric within the disk
with stronger emission in the NE, like the HI,
although the asymmetry is not as strong in H$\alpha$ as in HI.
While overall the H$\alpha$ disk emission is stronger in the NE, 
a luminous HII complex forming a striking 500 pc diameter bubble 
is located near the gas truncation radius in the SW.
There are HII regions associated with each of the 3 major extraplanar 
HI peaks.
H$\alpha$ is detected out to 40$''\simeq$3 kpc (projected) from the disk plane,
not quite as far out as HI is detected.
The extraplanar H$\alpha$ emission is also asymmetric like the HI, with 
the HII regions on the SW side being more luminous than those in the NE.

The radio continuum emission is fully discussed in Vollmer etal (2004).
Here we briefly summarize how the distribution of the 20cm emission
compares with the HI emission.
The radio continuum peak agrees well (within 1$''$) with the HI kinematic center,
and the radio continuum emission is more centrally concentrated than the HI,
as it is in most galaxies.
It does not show a significant asymmetry within the disk, as do the HI and H$\alpha$.
The radio continuum map
shows extraplanar emission, broadly consistent with the extraplanar HI distribution,
although the peak of extraplanar radio continuum is a bit inside the 
peak of extraplanar HI.


The only existing molecular observations of NGC 4522 are a 5 point CO (1-0)
mapping with the NRAO 12m telescope with a 55" beam (Smith and Madden 1997).
They measure a total H$_2$ mass of 3.2$\times$10$^8$ M$_{\rm sun}$ if the 
standard CO-H$_2$ relation applies. 
Thus in the central regions of NGC 4522
there are comparable amounts of HI and  H$_2$. Unfortunately there are
no CO data for the outer disk of NGC 4522. However, in the central regions
there is a hint that the CO may be distributed like the HI. There is
more gas to the North  and West and less East and South of the center.
Since H$_2$ is generally more centrally concentrated than HI in galaxies,
the ratio of extraplanar to disk HI+H$_2$ is probably less than the ratio 
of extraplanar to disk HI.

NGC~4522 has only $\sim$25\% of the HI expected for a galaxy of its mass,
so $\sim$75\% , or 1.2$\times$10$^9$ M$_{\sun}$ of HI has presumably been
stripped, heated, and ionized, perhaps by thermal evaporation (Cowie \& Songalia 1977;
Bureau \& Carignan 2002). The location and distribution of this gas is not known,
but could in principle be traced by its X-ray emission.
The ROSAT map shows weak extended X-ray emission at the 
projected location of NGC~4522 (B\"{o}hringer \etal 1994), 
although NGC~4522 itself has not been detected as
a localized source of X-ray emission
(Fabbiano \etal 1992; B\"{o}hringer \etal 1994; Snowden 1997).
If this heated gas could be detected, it would 
provide evidence about the interaction history of NGC~4522.

\section {Discussion}

NGC~4522 is a clear case of ISM gas selectively removed from
an undisturbed stellar disk.
The ISM removal from the outer disk, and the existence of HI only
on one side of the disk, 
argue strongly that the ISM has been pushed out of the disk by
ICM-ISM pressure, as the galaxy moves through the gas of the
Virgo intracluster medium. 
There is evidence that the galaxy is presently experiencing
ram pressure (this paper; Vollmer etal 2004), but it is not clear whether 
present pressure is strong enough to produce the observed stripping,
or whether is it in a post-peak pressure recovery phase with gas infall, 
as suggested by Vollmer etal (2000, 2001).
As explained below, the ram pressure required to produce the observed stripping
is about an order of magnitude larger than 
the ram pressure calculated according to standard assumptions.
Thus either NGC~4522 is in a post-peak pressure phase with gas infall,
or it is in an active phase and the  ram pressure is significantly higher
than the ``standard value''.

\subsection {Key Parameters of the ICM-ISM Interaction in NGC~4522}

There are several recent simulations which show what may happen to spiral galaxies 
in ICM-ISM interactions (Abadi \etal 1999; Quilis \etal 2000;
Vollmer etal 2001; Schulz \& Struck 2001).
While it cannot predict the detailed behavior of gas,
a simple ram pressure formula indicates the important parameters,
and the simulations indicate this formula 
yields a good first-order estimate of whether gas is stripped.
Gunn \& Gott (1972) proposed that
gas from a galaxy disk is stripped if the ram pressure $\rho$$_{\rm ICM}$v$_{\rm ICM}^2$ 
exceeds the gravitational restoring force per unit area $\frac{d\phi }{dz}\sigma _{\rm gas}$,
for a face-on encounter.
The resistance to stripping depends on galaxy properties including
the ISM gas surface density $\sigma _{\rm gas}$,
the gravitational force per unit mass $\frac{d\phi }{dz}$ =2$\pi$G$\sigma _{\rm disk}$,
due to the vertical z-gradient in the gravitational potential $\phi$,
and the disk mass surface density $\sigma _{\rm disk}$, which is dominated by stars.
The ram pressure from the ICM wind depends on
the ICM density $\rho$$_{\rm ICM}$,
the velocity of the galaxy with respect to the ICM v$_{\rm ICM}$,
and the encounter angle i between the disk and the ICM wind direction.
Simulations suggest stripping rates which vary approximately as cos$^2$i
(Vollmer \etal 2001 ; Schulz \& Struck 2001).
The ICM wind varies as a galaxy follows its orbital path within the
cluster, and the present state of a galaxy depends on this 
orbital history. Of particular importance is the time since (until)
maximum ram pressure.
Since many HI-deficient galaxies in clusters have highly radial
orbits (Dressler 1986; Solanes \etal 2001), 
the ram pressure can vary by factors of 20-100
during the orbit (Vollmer etal 2001).

Gas pushed outwards from the local disk will not necessarily be stripped from the
galaxy, since the global gravitational potential (including bulge, disk, and dark halo),
is deeper than that from the disk alone.
Gas pushed outwards from the disk may hang up in the halo, 
and some of it might fall back into the disk (Schulz \& Struck 2001),
especially in the case of a time variable ram pressure (Vollmer etal 2001).
A different approximation for stripping gas from the entire galaxy is 
that the ram pressure must exceed the force per area from the
radial gradient in the gravitational potential $\frac{d\phi }{dR}$=v$_{\rm rot}$$^2$/R,
which includes contributions from all the mass components of the inner galaxy,
not only the local disk.
This is only an approximation, since for a face-on encounter
the ram pressure is in the vertical direction, whereas
the gravitational acceleration v$_{\rm rot}$$^2$/R is in the radial direction,
so it does not strictly correspond to a restoring force.
Moreover, there is no guarantee that this pressure will accelerate gas
to the escape speed v$_{\rm esc}$=2$|$$\phi$$|$, which is at least $\sqrt{2}$ times larger
than the circular speed v$_{\rm rot}$ = $\sqrt{R\frac{d\phi }{dR}}$,
and includes contributions from mass outside the radius R.
So the Gunn \& Gott equation is a mininal stripping condition, 
but may be most appropriate for the extraplanar HI in NGC~4522, 
which we know has been removed from the disk but may not be escaping.

The 2 approximations yield similar values for the resistance of NGC~4522 to stripping.
The R band surface brightness of NGC~4522 at R=3 kpc,
corrected to face-on, is 22.8 mag arcsec$^{-2}$ (Koopmann etal 2001).
If we adopt a modest extinction correction of 0.3 mag, since
the outer disk has relatively little gas and dust, then 
the surface brightness at 3 kpc is 25 L$_{\sun}$ pc $^{-2}$,
and the stellar mass surface density is 100 M$_{\sun}$ pc$^{-2}$,
assuming a mass-to-light ratio of 4.
This yields a gravitational force per unit mass in the vertical direction
of $\frac{d\phi }{dz}$=9$\times$10$^9$ cm s$^{-2}$.
This compares with 
a gravitational force per unit mass in the radial direction of
of $\frac{d\phi }{dR}$=11$\times$10$^9$ cm s$^{-2}$,
assuming a rotation speed of 
v$_{\rm rot}$=103 km s$^{-1}$ at R=3 kpc (Rubin \etal 1999).

Is the present ram pressure on NGC~4522 sufficient to strip it?
We first assume a standard, smooth, static ICM,
with a density n$_{\rm ICM}$=$\rho$$_{\rm ICM}$/m$_{\rm p}$=10$^{-4}$ cm$^{-3}$ (\S 5.2),
and a typical Virgo galaxy velocity  of v$_{\rm ICM}$= 1500 $\kms$,
and a typical gas surface density of $\sigma _{\rm gas}$=10 M$_{\sun}$ pc$^{-2}$,
and an encounter angle of 45$\deg$.
With these ``standard'' values, which are discussed below,
the acceleration due to ram pressure $\rho$$_{\rm ICM}$v$_{\rm ICM}^2$cos$^2$i/$\sigma _{\rm gas}$
is 1$\times$10$^9$ cm s$^{-2}$,
an order of magnitude less than needed to strip the gas at R=3 kpc in NGC~4522.
Either the galaxy must presently be in a post-peak pressure phase,
with the extraplanar gas now possibly falling back into the galaxy (Vollmer etal 2000),
or the ram pressure must be greater than that predicted from
the nominal Virgo ICM parameters.
Here we discuss the nominal Virgo ICM parameters,
and in the next section we discuss the possibility that the ram pressure 
presently experienced by NGC~4522 is significantly higher than the ``standard value''.



NGC~4522's radial velocity of 2337 $\kms$ is on the high end of the 
distribution for Virgo galaxies, implying that it has a
velocity of at least 1300 $\kms$ with respect to the mean cluster velocity
of 1050 $\kms$ (Binggeli, Popescu, \& Tammann 1993).
There must also be a significant velocity component in the plane of the sky 
to cause the extraplanar filaments in this nearly edge-on galaxy,
although it is unlikely that this component is as large as
the extreme l-o-s component.
If the plane-of-sky component is 700 $\kms$, then the total velocity is 1500 $\kms$,
a typical orbital velocity for a Virgo cluster galaxy.
If the plane-of-sky velocity component is much larger,
the galaxy would not be bound to the cluster.
Although some cluster galaxies could be given 
large orbital speeds through tidal interactions in merging subclusters,
it is highly unlikely that  NGC~4522 has an orbital
speed as high as the 4000 km s$^{-1}$ needed to account for the observed stripping.

The determination of the plane-of-sky ICM motion and the disk encounter angle
are coupled, with
smaller plane-of-sky ICM motions corresponding to more edge-on encounters.
The encounter angle in NGC~4522 is likely in the 30-60$\deg$ range,
since its large line-of-sight velocity and edge-on orientation insure that the
encounter angle cannot be too close to face-on,
and the extraplanar filaments insure that it cannot be too close to edge-on.
Simulations  at various encounter angles i 
by Vollmer \etal (2001) find stripping rates which vary 
approximately as cos$^2$i, except that for edge-on encounters 
the stripping rate is a few percent of the face-on case, rather than zero.
Simulations by Schulz \& Struck (2001) show that
while the initial (first 200 Myr$\sim$1 dynamical time) stripping rates do vary approximately 
as cos$^2$i (for i=0-60$\deg$),
the total stripping rates over longer times (500 Myr) have a much weaker i dependence,
perhaps because viscosity effects become more important than ram pressure on
longer timescales (Abadi etal 2001).
Thus the pressure requirements are minimized for a face-on encounter,
and 2 times higher for an encounter angle of 45$\deg$.


While NGC~4522 could be in a post-peak pressure phase,
its orbit is not a simple radial orbit.
It has an extreme  line-of-sight velocity in the cluster,
yet its projected location is not close to the center, as it would be if it were on a highly 
radial orbit bound to the cluster (Vollmer etal 2001).
NGC~4522 is one of the galaxies which gives the spirals in Virgo a multi-peaked,
non-Gaussian velocity distribution, which is
likely due to infalling galaxies, groups, and subclusters (Huchra 1985;
Schindler etal 1999; Conselice etal 2001; Vollmer etal 2001).
The galaxies in such systems have not yet reached dynamical equilibrium
in the cluster.
This is consistent with NGC~4522 being a recent arrival in the cluster.
The radio continuum maps show enhanced polarization 
along the eastern edge (Vollmer etal 2004), 
suggesting that the galaxy's plane-of-sky motion with respect to the ICM
is more toward the East than the south, as it would be for a highly radial orbit around M87.
It could be on an orbit whose closest approach to the cluster center brought it
to the west of M87 on the near side of the cluster.

\subsection{Strong Ram Pressure in a Dynamic ICM?}

If the present ram pressure is causing the observed stripping in NGC~4522,
the ram pressure must be stronger than suggested by standard assumptions.
The ICM gas density could be higher if there are gas lumps or cool gas,
and the ICM wind velocity could be higher
if NGC~4522 has an unbound, high-velocity orbit or if the Virgo cluster
has bulk ICM motions.

Recent observations (Dupke \& Bregman 2001)
and simulations  (e.g., Ricker 1998; Takizawa 2000; Roettiger \& Flores 2000)
have shown that the ICM in many clusters,
rather than being smooth and static,
is instead shock-filled and dynamic, with bulk speeds of 1-few 1000 km s$^{-1}$,
due to ongoing subcluster mergers.
This can cause significantly higher ram pressure for some of the cluster galaxies.

The ICM gas density is generally estimated from
azimuthally-averaged radial surface brightness profiles
of X-ray emission and gas temperature, 
and the assumption of a locally smooth gas distribution.
These assumptions lead to an estimated ICM density of n$\sim$5$\times$10$^{-5}$ cm$^{-3}$ 
at a distance of 900 kpc from M87, which corresponds to 3.3$\deg$ (Fabricant \& Gorenstein 1983).
However, the ICM is not radially symmetric in Virgo, and there are concentrations of X-ray emission
around the giant ellipticals M87, M86, and M49, as well as in a ridge extending between M87 and M49
(B\"{o}hringer \etal 1994) where NGC~4522 is located (Figure 16).
ASCA X-ray spectroscopy show that ICM gas at
the western side of this ridge  has unusually high temperatures,
while the eastern side has temperatures similar to the rest of the cluster 
(Kikuchi etal 2000; Shibata etal 2001).
Thus the surface brightness enhancement of this ridge is due to both a density enhancement
and a temperature enhancement, which are themselves probably due
to shocks caused by the infall of the M49 sub-cluster toward M87. 
The estimated gas density in this ridge region
is n$\sim$10$^{-4}$ cm$^{-3}$ (Kikuchi etal 2000), which is perhaps a factor of 2 higher 
than the average value at this distance from M87. 
Thus NGC~4522 may have recently passed through a region of higher than average
ICM density in Virgo.
The HI morphology and radio continuum maps suggest
that NGC~4522's plane-of-sky motion is toward the east or southeast, 
and Figure 16 shows that this direction is roughly
away from the ridge of X-ray emission between M49 and M87.
The motions of the ICM in Virgo are not known, but the region between
M49 and M87, with evidence for ICM shocks, could have large-scale ICM motions,
associated with the M49 sub-cluster merger (Schindler etal 1999; Shibata etal 2001).
Bulk ICM motions which are comparable to, or larger than,
those of the galaxies with respect to the cluster (Dupke \& Bregman 2001),
can increase the relative velocity between some galaxies and the ICM by a factor of
$\sim$2. 
Shocks can cause local enhancements in gas densities of factors of $\sim$2-4. 
Together, these dynamic ICM effects can increase 
the ram pressure P$_{\rm ram}$$\sim$$\rho$$_{\rm ICM}$v$_{\rm ICM}^2$ 
by as much an order of magnitude for some cluster galaxies.
This would be enough to cause the observed stripping in NGC~4522.

It is also possible that the mean ICM density is higher than that indicated
by the X-ray emission, if there are significant amounts of cooler, 10$^5$-10$^7$ K gas
in the cluster. 
Outside of cluster cores,
most of the baryons are believed to be in intergalactic
gas at these temperatures (Cen \& Ostriker 1999; Mathur, Weinberg, \& Chen 2003),
and there may be signficant amounts of 10$^5$-10$^7$ K gas in at least
the outskirts of clusters.

%



\subsection {HI Morphological and Kinematic Asymmetries as Evidence for Ongoing ICM Pressure}

There are several morphological and kinematic asymmetries in NGC~4522
which provide important clues about its interaction with the ICM.
In most cluster spirals, 
the disk encounters the ICM wind at an intermediate angle between face-on and edge-on,
and this should produce asymmetries in the galaxy,
both because there are leading and trailing sides to the disk,
and also because one side is rotating into the ICM.
As explained below, some of the asymmetries in NGC~4522 
are naturally understood as the consequences of ICM pressure plus rotation.

The HI asymmetry in the disk, with relatively
weak disk HI emission and a high ratio of
extraplanar to disk emission for the SW side,
could be naturally explained by gas being 
preferentially stripped on the SW side.
The SW half of the disk is rotating into the oncoming ICM,
so the ICM pressure should be highest on the SW side,
and significantly less on the NE side.
Assuming a galaxy velocity through the ICM of 1500 km s$^{-1}$,
and a rotation velocity of 100 km s$^{-1}$, the SW side which is rotating into
the oncoming ICM, should experience a ram pressure (1600/1400)$^2$=1.3 times stronger than
the NE side, which is rotating away from the oncoming ICM.
This difference is less if the galaxy-ICM velocity is significantly higher.

The broad lines and the blueshifts
could be natually explained by significant ongoing ICM pressure, which would
push the ISM gas to lower, closer to cluster systemic, velocities.

\subsection{ Comparison with Simulations}

Is the galaxy experiencing strong ICM pressure now,
or is it in a recovery phase with gas infall, as
suggested by Vollmer etal (2000)?
We have compared the HI morphology and kinematics
with those of the N-body, sticky particle simulations
of Vollmer etal (2000), who attempted to match the H$\alpha$ data
on NGC~4522 available at that time. 
In their favored simulation, the peak ram pressure occurred 650 Myr in the past,
and the simulated galaxy is presently experiencing gas infall.

There are both similarities and differences between the data and the 
Vollmer etal (2000) simulation.
They have roughly similar radial and z extents of disk and extraplanar HI gas,
and in both the NE extraplanar peak is weaker than the SW one.
Both exhibit an 
HI surface brightness asymmetry in the disk, although the details differs.
The significance of the differences is limited because the
simulations show all the gas, whereas the HI data shows only one component of the ISM.

There are also several features in the data which were not reproduced in the
simulations.
One key difference is that the amount of extraplanar gas is larger in the data  than in the
fall-back model.
In the data, 40\% of the HI is extraplanar, and
the peak HI surface density in the extraplanar component is 67\% of the peak in the disk.
If all the molecular gas is assumed to be in the disk,
then the corresponding values for the total gas mass distribution  would be $\sim$25\% and 
less than  33\% , respectively.
In the Vollmer etal (2000) model,
15\% of the gas is extraplanar,
and the peak gas surface density in the extraplanar component is 10\% of the peak in the disk.
Large quantities of extraplanar gas located close to the truncation radius
in the disk are generally seen only in early phases of stripping
(Vollmer \etal 2001; Schulz \& Struck 2001), and not during fall-back phases
(Vollmer etal 2000; 2001). Since the angular momentum of the gas is not conserved,
gas pushed outwards by ram pressure will generally not all return to the disk positions
from which they originated.
Another difference in gas morphology concerns
the existence in the data of an extraplanar HI surface density peak in the NE 
as well as the SW,
each at a larger radius than the outermost disk emission, and
both at the same height above the disk plane.
This NE peak does not have a counterpart in the model.

There are also significant differences in the kinematics.
In the data, much of the extraplanar gas 
exhibits a blueshift with respect to the nearby disk emission.
In particular, the SW extraplanar component is a bit blueshifted in the data,
but the model shows similar or even redshifted
velocities with respect to nearby disk emission.
Furthermore, in the data, some of the extraplanar gas has large linewidths,
with blueshifted tails, whereas
in the model, the extraplanar gas has fairly symmetric spectra and 
narrower linewidths.
The broad lines and the blueshifts
could be due to significant ongoing ICM pressure.

In summary, the Vollmer etal (2000) simulations of NGC~4522 in a post-active stripping phase
provide an approximate match to the 
observed projected radial and vertical extent of gas,
but do not match the amount of extraplanar HI, the disk asymmetry,
or the kinematics.
New simulations by Vollmer etal (in prep) of NGC~4522 in an active stripping phase
with a ram pressure an order of magnitude higher than the ``standard value''
shows a much better match to all these properties.

\subsection {How NGC~4522 Compares with Other Galaxies 
With Evidence for ICM-ISM Interactions}

In the nearby Virgo cluster, there are a few other galaxies with 
good evidence for ongoing  ICM-ISM interactions,
including NGC~4654,  NGC~4548, NGC~4388, and NGC~4569
(Phookun \& Mundy 1995; Vollmer 2003;  Vollmer etal 2001b; 
Yoshida etal 2002; Vollmer \& Hutchmeier 2003;
Tschoke etal 2001; Kenney etal 2004).
These galaxies differ from NGC~4522 in the surface density 
of extraplanar HI, and in the extraplanar gas morphology.

None of the other galaxies have a surface density of extraplanar HI
as high as NGC~4522. 
Simulations generally show a high 
surface density of extraplanar gas only in the early phases
of an ICM-ISM interaction (Vollmer etal 2001; Schulz \& Struck 2001).
In each of  NGC~4548, NGC~4654, and NGC~4569,
one extraplanar or outer disk gas arm
emerges from the edge of a truncated gas disk. 
Such an ISM morphology resembles those in later phases of the simulations of 
Vollmer etal (2001) and Schulz \& Struck (2001),
in which a combination of rotation plus ICM pressure produces a
dominant gas arm in the outer galaxy.
The extraplanar gas in NGC~4522, however, does not 
have a simple spiral arm morphology, but more of a bow-shock
type morphology. In simulations, this type of morphology is seen in 
earlier phases of the ICM-ISM interaction, before the arm morphology
has time to develop.
Thus the differences in extraplanar gas density and morphology
are both consistent with  NGC~4522 being in an earlier phase of
an ICM-ISM interaction than the other galaxies.

\subsection {Any Evidence for a Shock?}

Is there direct evidence for a shock, or strong ICM pressure currently on NGC~4522?
There is enhanced polarized radio continuum emission on the eastern edge of the disk,
on the opposite side from the extended extraplanar HI (Vollmer etal 2004).
This indicates a compressed magnetic field and probably enhanced ICM pressure on this side.
The same general region also has the flattest spectral index, indicating local
acceleration of relativistic electrons.
This suggests the existence of a shockfront.

There is presently no optical or X-ray evidence for a shock at the ICM-ISM interface.
Deep H$\alpha$ images do not show any diffuse emission outside the main body of the galaxy
which might be associated with a shockfront (Kenney \& Koopmann 1999),
and NGC~4522 has not been detected as a localized source of X-ray emission (\S 4.3).
The X-ray emission from an ICM-ISM interaction is generally expected to be 
modest compared with the ICM X-ray background,
since cluster galaxies are not moving highly supersonically in the ICM,
with expected Mach numbers of $\sim$1-5 (Stevens etal 1999; Veilleux etal 1999).
Stevens etal (1999) simulate an ICM-ISM interaction for a smooth ISM in a spherical galaxy, 
which differs from the actual, lumpy, star-forming disk
ISM of a spiral galaxy, but their simulations provide a first-order estimate of
the X-ray emission which might be expected in NGC~4522.
Model 2c of Stevens etal (1999), which is the case most similar to NGC~4522 in Virgo, 
predicts L$_{\rm x}$$\sim$10$^{40}$ erg s$^{-1}$, assuming an emission region with 5 kpc radius.
This is below the level observed by Einstein, L$_{\rm x}$$<$1.9$\times$10$^{40}$ erg s$^{-1}$
(Fabbiano etal 1992), but should be detectable with Chandra or XMM.
Most of this X-ray emission is predicted to come from a tail of stripped gas,
with much weaker emission arising from the shock front. 
Thus the lack of observed X-ray emission in NGC~4522
is not inconsistent with ongoing ICM pressure.
More sensitive X-ray observations may provide a more definitive answer,
although the weakness of the shock emission, as well as projection effects
and confusion with disk emission,
may make it difficult to clearly identify a shockfront.

\subsection{The Fate of the Gas: Is Molecular Gas Stripped?}

The response of a turbulent, multi-phase ISM to ram pressure is difficult to assess,
but a key question is the fate of molecular clouds, whose high surface densities
of $\sim$200 M$_{\sun}$ pc$^{-2}$ (Larson 1981; Scoville etal 1987; Solomon etal 1987)
make them difficult to strip by ram pressure (Kenney \& Young 1989).
The similarity of the HI and H$\alpha$ distributions in NGC~4522 (\S 4.3) 
is significant because it
implies that the star-forming ISM has been effectively removed from the outer galaxy.
While the HII regions in the extraplanar gas could have formed within molecular clouds
formed in situ from gas that had lower densities when it was stripped, the
absence of HII regions beyond the HI truncation radius
indicates that even molecular gas has been effectively removed from the
outer disk of NGC~4522.


The present result indicates that at least for NGC~4522, the outer disk has been 
effectively stripped of all its gas. 
Previous single dish CO results indicate that HI deficient cluster galaxies have
nearly normal CO luminosities (Kenney \& Young 1989; Boselli etal 1997).
Considering the very different radial distributions of
molecular and atomic gas in galaxies,
with most HI in the outer parts of galaxies and most molecular gas 
in the inner parts (e.g. Kenney \& Young 1989; Young \& Scoville 1991)
it could be true that in general that HI deficient galaxies have virtually all the
gas removed from their outer disks.
Thus low resolution observations
which measure the global CO luminosity are not that sensitive to outer disk
molecular gas stripping.
This, together with the fact that
uncertainties in CO deficiency are higher than those for HI deficiency
(Boselli etal 1997)
makes it hard to know whether strongly HI-deficient galaxies might be slightly 
H$_2$-deficient.

In many Sc galaxies the CO radial distribution resembles an exponential with a scale 
length similar to the optical starlight (Young \& Scoville 1991). 
The HI stripping radius of 40$''$ is 1.33 times NGC~4522's R-band exponential 
scale length of 30$''$ (Koopmann etal 2001).
Since the central  1.33 scale lengths of an exponential disk contains 38\% of the total flux,
a galaxy completely stripped beyond 1.33 scale lengths would have
a (logarithmic) CO deficiency parameter of 0.4. Since NGC~4522 likely contains 
extraplanar molecular gas
in addition to inner disk gas, its expected CO deficiency would be somewhat less than this.
The actual CO deficiency parameter for NGC~4522 is 0.2, corresponding to 63\% of the average
CO luminosity for non-cluster spiral galaxies with the same optical diameter,
according to the formulae in Boselli etal (1997).
This is consistent with both the expected deficiency and with normalcy,
given the large standard deviation of 0.3 (in logarithmic units)
for the CO luminosity of non-cluster spiral galaxies of a given optical size.

Even if molecular clouds are too dense to be directly and individually stripped,
effective stripping of molecular clouds might occur in 3 ways.
1. GMCs quickly evolve to a lower density state through the natural star formation-ISM cycle, 
at which point they are stripped by ram pressure.
The lifetimes of molecular clouds are estimated to be $\leq$10$^7$ yrs 
(Larson 2003; Hartmann 2003),
significantly shorter than the orbital and stripping timescales.
2. The low density ISM is stripped from around the denser molecular clouds, 
which are initially left behind, but are then ablated by hydrodynamic 
(Nulsen 1982; Quilis etal 2000) or other effects. 
3. Molecular clouds might be tied to the rest of the ISM, e.g. by magnetic fields, so that
the entire ISM is stripped out together. 

Whereas molecular gas in the outer disks of cluster spirals may be effectively stripped,
inner galaxy regions could be different, although observationally the situation is unclear.
The Coma cluster spiral NGC~4848 is a possible although not clearcut case 
for low density HI stripped from the central part of a galaxy,
while leaving, at least temporarily, the molecular gas behind (Vollmer etal 2001).
There is also suggestive evidence for high CO/HI ratios and HI modest deficiencies in 
the central regions of globally HI-deficient Virgo galaxies (Kenney \& Young 1989),
although because the CO/HI ratio depends strongly on internal conditions in the ISM,
it is not clear that any gas has actually been lost from inner disks.
The fate of molecular gas in stripping may depend on factors such as the local H$_2$/HI ratio,
and it may be easier to get effective stripping of molecular clouds if the 
local ISM is HI-dominated.
The global M$_{\rm HI}$/M$_{\rm H_2}$ ratio in NGC~4522 is about 1,
and at the gas truncation radius of 3 kpc, the amount of molecular gas is probably 
less than the amount of HI.

Nearly normal CO luminosities in HI deficient cluster spirals had been
previously proposed to mitigate the impact of HI gas stripping on galaxy evolution
(Kenney \& Young 1989; Boselli etal 1997). The present results indicate that 
molecular clouds do not prevent outer disks from being completely stripped of their ISM,
although they could still help the inner disk ISM survive.

\subsection{ The Fate of the Gas: Ram Pressure Stripping vs. Star Formation}

Recent simulations show that in significant ram pressure stripping events,
most of the gas is stripped relatively quickly, in less than 10$^8$ yrs 
(Abadi \etal 1999; Quilis \etal 2000;
Vollmer etal 2001; Schulz \& Struck 2001), although significant 
stripping at lower rates may continue for a much longer time.
NGC~4522 has lost over 10$^9$ M$_{\sun}$ of gas, so the peak
stripping rate is $\sim$10 M$_{\sun}$ yr$^{-1}$.

The total star formation rate in the galaxy,
based on the H$\alpha$ luminosity with no extinction correction,
is 0.1 M$_{\sun}$ yr$^{-1}$ (Kenney \& Koopmann 1999).
Since the galaxy is highly inclined, there must be a significant
extinction correction, but with a global ratio of
far-infrared luminosity to optical luminosity L$_{\rm FIR}$/L$_{\rm opt}$ of 0.5,
the correction is unlikely to be more than a factor of a few.
The rate of star formation triggered by the interaction is 
an unknown fraction of the total star formation rate.
Thus the peak gas stripping rate is $\sim$2 orders of magnitude greater
than the present rate of ICM-triggered star formation.
Unless the peak star formation rate is much larger than the present
star formation rate, gas loss due to ICM pressure strongly dominates
triggered star formation due to ICM pressure.

Although the central star formation rate normalized by the
optical luminosity is somewhat higher for NGC~4522 than the average for
Virgo cluster and isolated galaxies,
it is within a factor of 2 of the median value for isolated spirals
(Kenney \& Koopmann 1999).
The H$\alpha$ equivalent width in the central 30\% of the disk corresponds to
a stellar birthrate parameter, defined as the ratio as the present to the
past average star formation rate, of 0.1-0.2, according to the precepts in
Kennicutt, Tamblyn, \& Congdon (1994) and Boselli etal (2001).
The lower value is based on the H$\alpha$ flux with no extinction correction,
and the upper value assumes a factor of 2 extinction correction.
In terms of long-term galactic evolution, the outer disk of the galaxy would probably 
be closer to models which have star formation terminated abruptly in time,
rather than a strong starburst followed by termination.
However, since the central gas disk will likely survive,
the global star formation rate will be reduced but not terminated,
and the long-term evolution of the galaxy
may resemble models in which the star formation 
is reduced over long timescales.

\subsection{The Importance of Subcluster Merging}

We have reached the somewhat surprising conclusion that despite
NGC 4522's location somewhat outside the cluster core at 0.8 Abell radii,
it is an excellent candidate for ongoing strong ram pressure stripping.
Quite possibly the ongoing merger between the M49 group and Virgo
proper has stirred up the ICM, locally enhancing the ram pressure.
This would imply that the impact and reach of the ICM is
closely related to the dynamical state of the cluster.
If this is indeed the case it may help explain a number of
phenomena seen in the outskirts of clusters (van Gorkom 2003), such
as the HI deficiency seen out to 2 Abell radii (Solanes et al 2001),
and suppressed star formation rates as far out as 2 R$_A$
(Balogh et al. 1998; Gomez et al. 2003; Lewis et al. 2002).

\section{ Conclusions}

1. The Virgo cluster Sc galaxy 
NGC~4522 is a clear case of ISM gas selectively removed from
an undisturbed stellar disk by an ICM-ISM interaction.
The existence of a one-sided distribution of extraplanar gas 
and an undisturbed stellar disk
clearly indicates a non-gravitational external process
which has selectively removed ISM gas.
HI is truncated within the disk at R=3 kpc = 0.4R$_{25}$.
40\% of the HI appears to be extraplanar, 
and all the extraplanar gas is displaced toward one side of the major axis,
extending 60$''$=5 kpc above the disk plane.
The peak surface density of the extraplanar HI is 10 M$_{\sun}$ pc$^{-2}$,
a high value for extraplanar gas.

2. There is direct evidence for ongoing ICM pressure on NGC~4522
from radio continuum maps (Vollmer etal 2004), but
it remains unclear whether the present ram pressure is
sufficiently strong to to have caused the observed removal of the disk gas,
or whether the galaxy may presently be observed some time after peak ICM pressure.
Both the kinematics and the morphology of the HI
are more suggestive of ongoing stripping than 
with gas fall-back which may occur long after peak pressure.

3. The HI content within the disk is asymmetric, with 50\% less emission
on the SW, high-velocity half of the disk. 
An opposite asymmetry is observed in the halo, which has more HI in the 
SW than the NE.
This could be due to the
stronger ICM pressure and enhanced stripping
which is expected in the half of the disk 
which is rotating into the oncoming ICM.

4. Much of 
the extraplanar HI exhibits a modest net blueshift with respect to the 
systemic velocity of the galaxy, consistent with stripped galaxy gas being
accelerated toward the mean cluster velocity.
Extraplanar gas on the  SW, high-velocity half of the disk
shows large linewidths (FWZI) of 150 km s$^{-1}$, including a tail of weaker emission
extending to lower velocities. We suggest that the lower velocity tails
represent gas which is accelerated toward the  mean cluster velocity.

5.
The galaxy is 3.3$\deg$$\simeq$800 kpc from M87, which is somewhat outside the
region of strongest X-ray emission, where the ICM gas density is highest.
Assuming a smooth and static ICM,
the calculated ram pressure at this location appears
inadequate by an order of magnitude to cause the observed stripping.
We consider the possibility that 
in a dynamic, shock-filled ICM with bulk motions and local density enhancements,
the ram pressure presently on NGC~4522 may be significantly stronger than it would
be in smooth, static ICM.




6.
NGC~4522 has only $\sim$25\% of the HI expected for a galaxy of its mass,
so $\sim$75\% , or 1.2$\times$10$^9$ M$_{\sun}$ of HI has presumably been
stripped, heated, and ionized. The location and distribution of this gas is not known,
but could in principle be traced by its X-ray emission, and if detected 
would provide evidence about the interaction history of NGC~4522.

7.
The inferred peak gas stripping rate of $\sim$10 M$_{\sun}$ yr$^{-1}$
is much larger than the galaxy's total star formation rate of 
$\sim$0.1 M$_{\sun}$ yr$^{-1}$,
implying that the present rate of triggered star formation due to ICM pressure
is minor compared to the peak rate of gas loss due to ICM pressure.

8.
The HI and H$\alpha$ distributions in NGC~4522 are similar. 
Both are truncated at R=3 kpc
in the disk, and they are coextensive in the extraplanar components.
This implies that the star-forming ISM, presumably including the molecular gas,
has been effectively removed from the galaxy beyond a radius of R=3 kpc.
If this result is general, it would indicate that
molecular gas does not strongly mitigate the effect of
stripping on cluster galaxy evolution. The outer disks of cluster spirals
can be stripped clean by ram pressure stripping. In clusters with a denser ICM
and stronger ram pressure, most of the disk could be completely stripped,
transforming spiral galaxies into lenticulars.

\section{ Acknowledgements}

We thank the VLA Director for granting time for these VLA observations,
Curt Struck for helpful discussions, and Sabine Schindler for providing the X-ray map.
This is reseach supported by NSF grants AST-0071251 to Yale University (JDK)
and AST-0098294 to Columbia University (JVG).
The National Radio Astronomy Observatory is a facility of the National Science Foundation 
operated under cooperative agreement by Associated Universities, Inc.

\begin{figure}
\epsscale{0.94}
\plotone{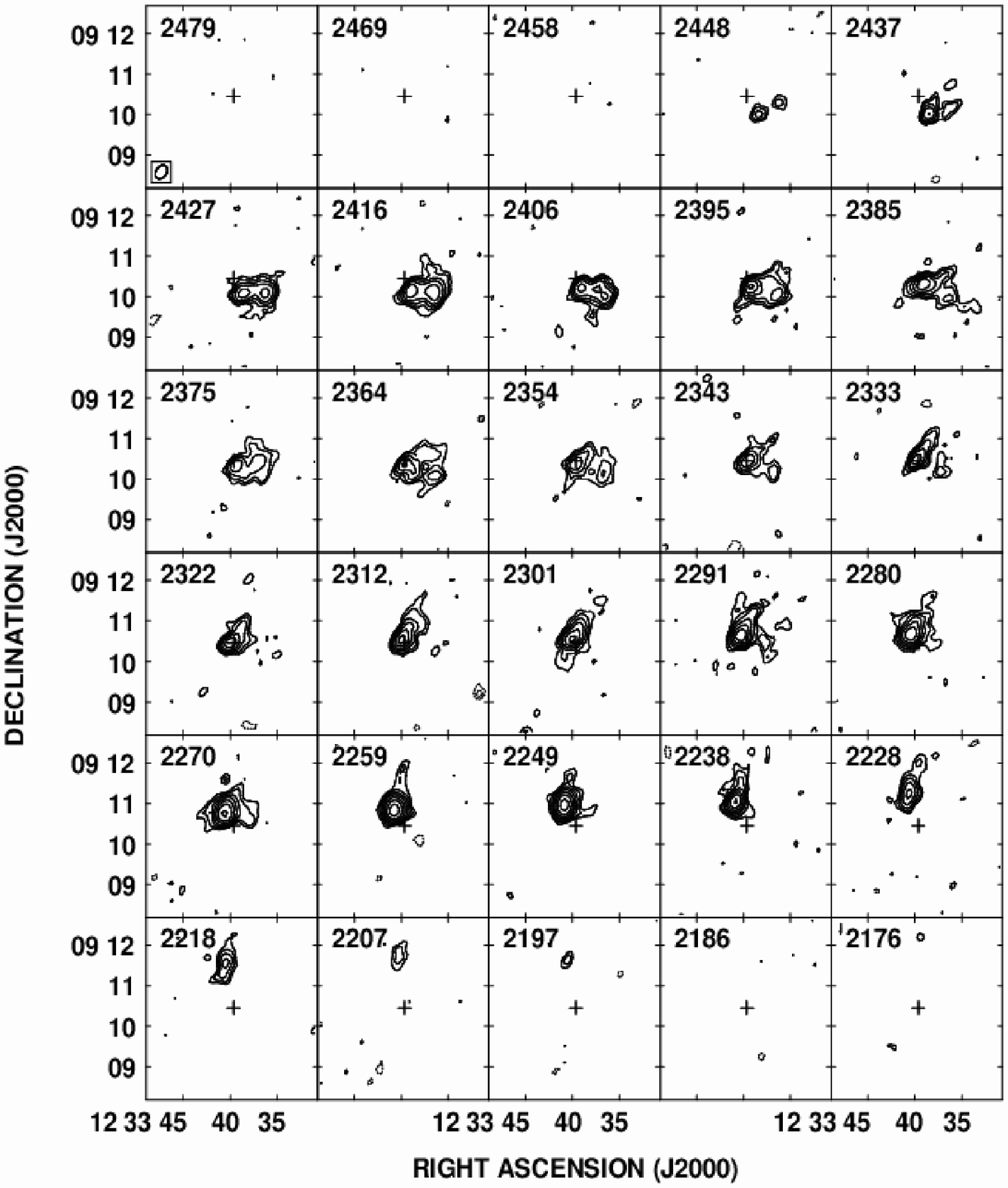}
\caption{
HI channel maps of NGC~4522. Velocity given in upper left
of each panel.
Lowest contour levels are $\pm$1.35 mJy/beam = 3$\sigma$,
and subsequent contour levels are in increments of factors of $\sqrt 2$.
Dashed contours are shown for negative values.
Cross marks HI kinematic center.
}
\end{figure}

\begin{figure}
\epsscale{0.95}
\plotone{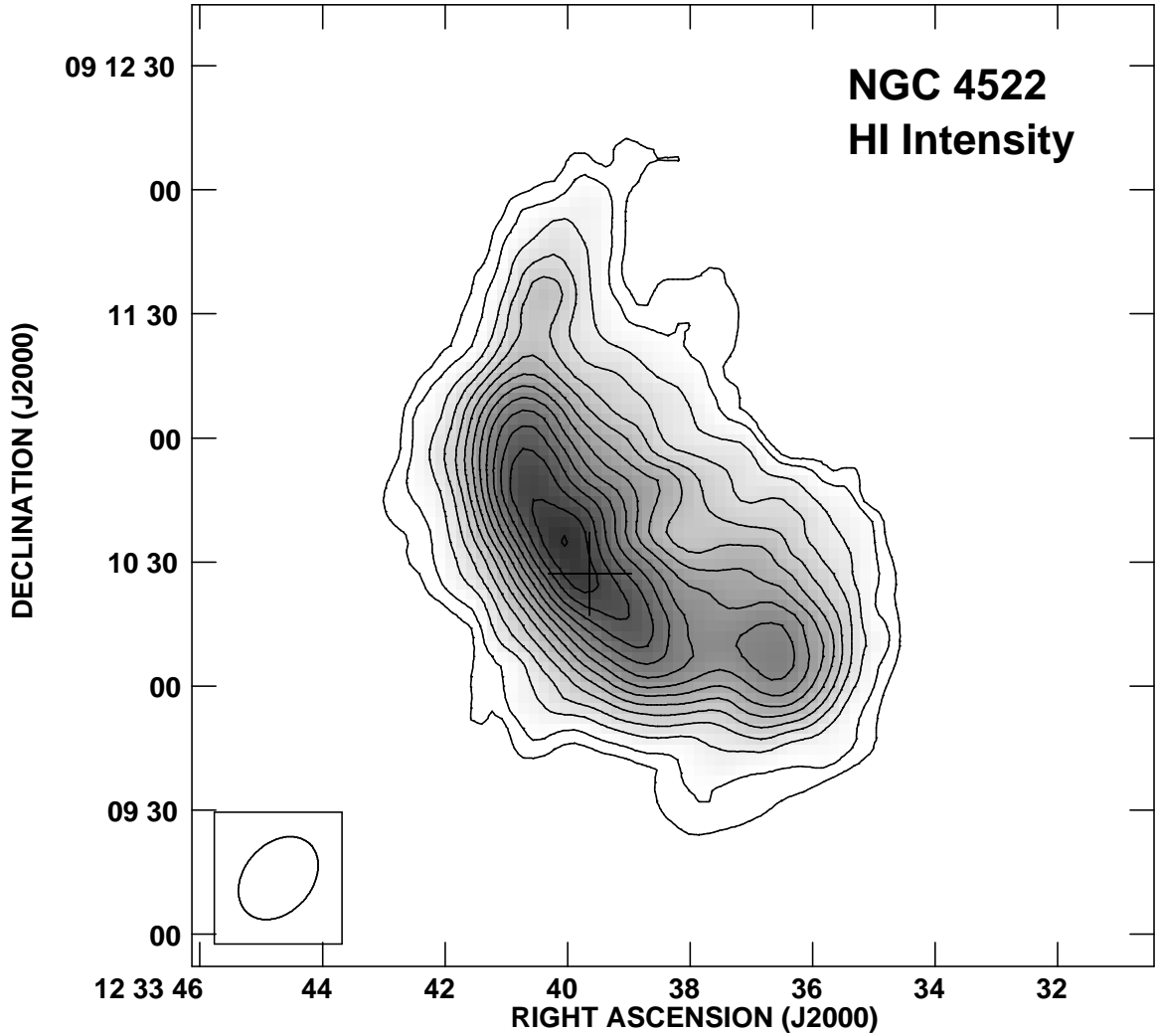}
\caption{
HI total intensity (moment 0)  map. Cross marks HI kinematic center.
Note peak in projected HI surface density to the NE of nucleus,
and the HI asymmetry in the disk.
Lowest contour is 25 mJy/beam km/s,
which corresponds to  0.6 M$_{\sun}$ pc$^{-2}$ = 8$\times$10$^{19}$ cm$^{-2}$.
Contour increments are 
0.8, 1.6, 3.2, 4.8, 6.4, 8.0, 9.6, 11.2, 12.8, 14.4, 16.0, 17.6, 19.2, 20.8$\times$10$^{20}$ cm$^{-2}$.
}
\end{figure}

\begin{figure}
\epsscale{0.95}
\plotone{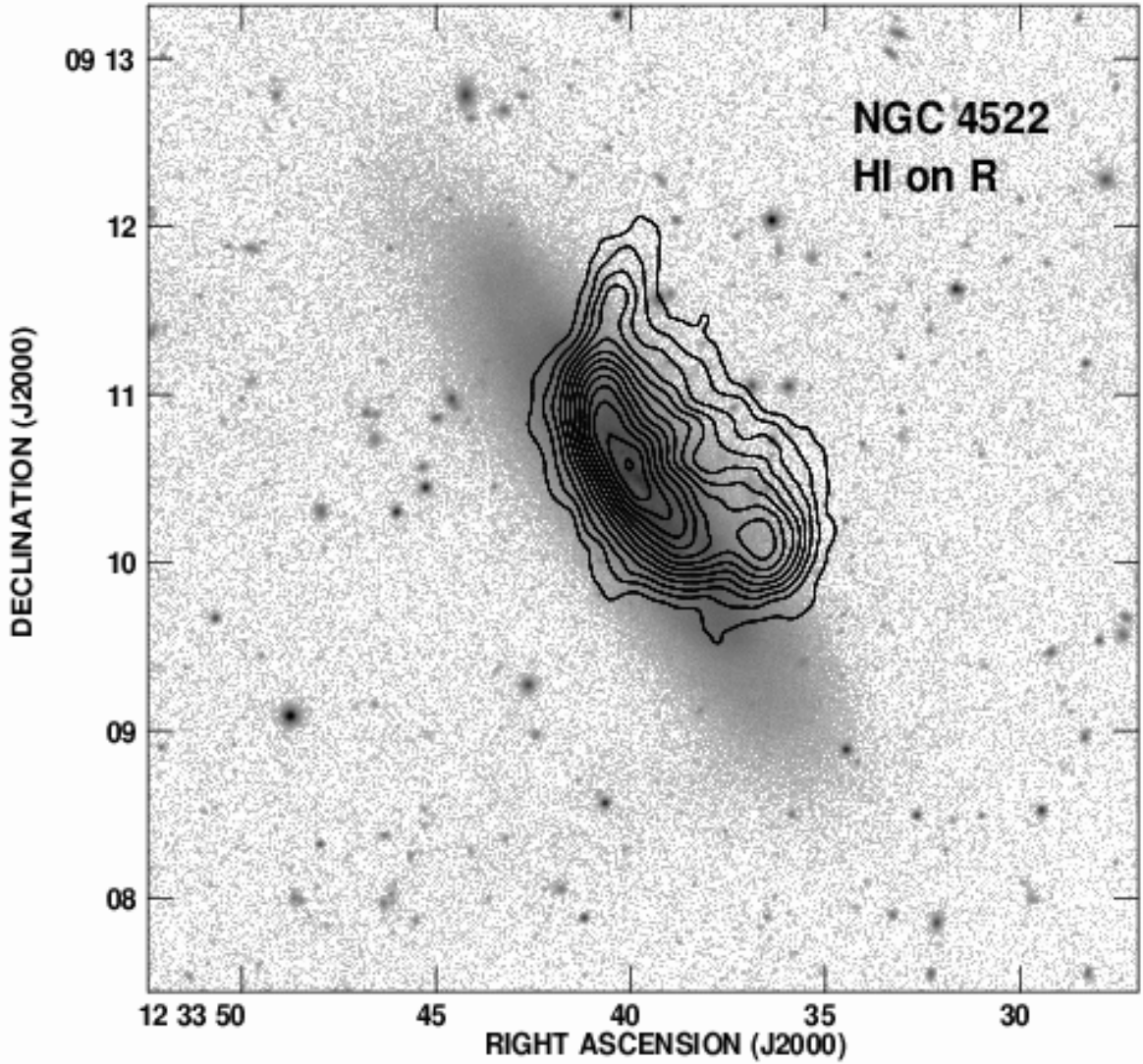}
\caption{HI moment 0 contour map  of NGC~4522 on R-band greyscale image
from the WIYN telescope from Kenney \& Koopmann (1999).
Lowest HI contour level and contour increments are 50 mJy/beam km/s.
Optical image shown with logarithmic stretch.
Note undisturbed outer stellar disk.
}
\end{figure}

\begin{figure}
\epsscale{0.95}
\plotone{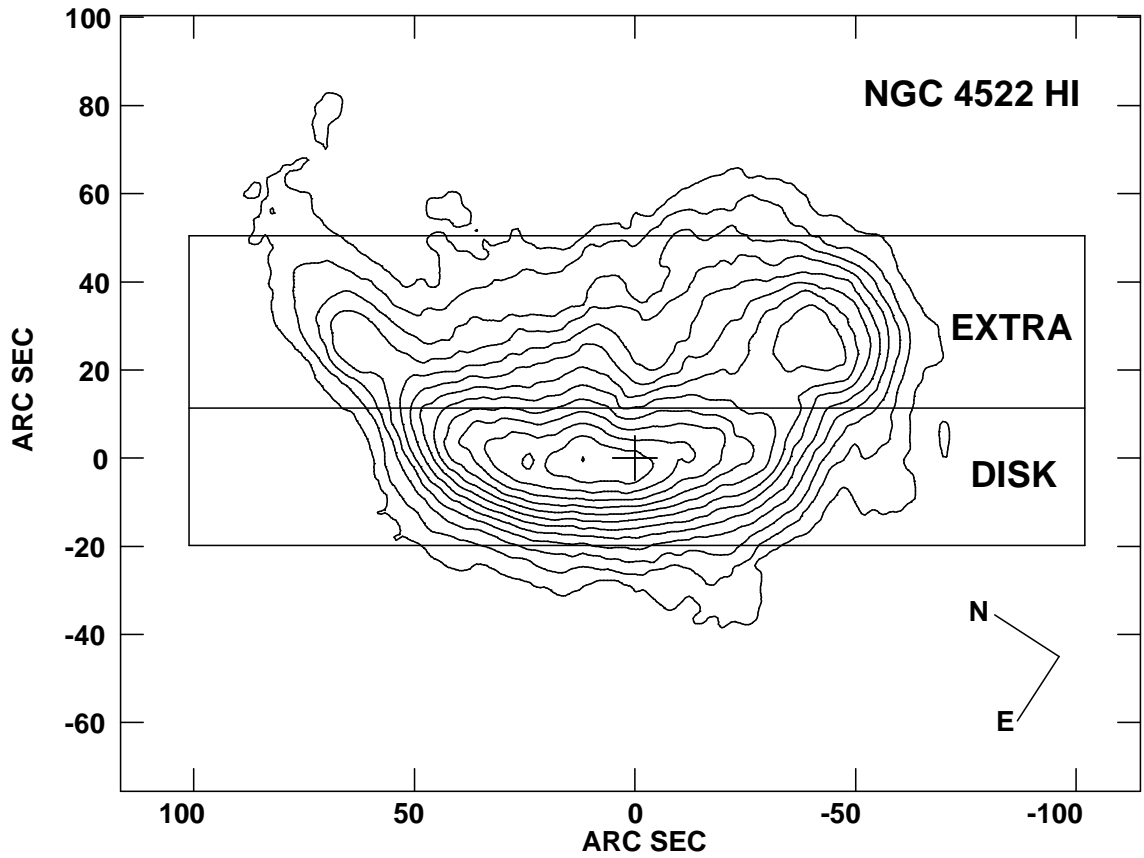}
\caption{HI moment 0 map of NGC~4522, rotated by 57 degrees so that
the major axis is oriented horizontally. The inset boxes marked
identify the regions sampled in the associated position-velocity plots
in Figure 10, which are from cuts parallel to the major axis of the galaxy.
}
\end{figure}

\begin{figure}
\centering
\plotone{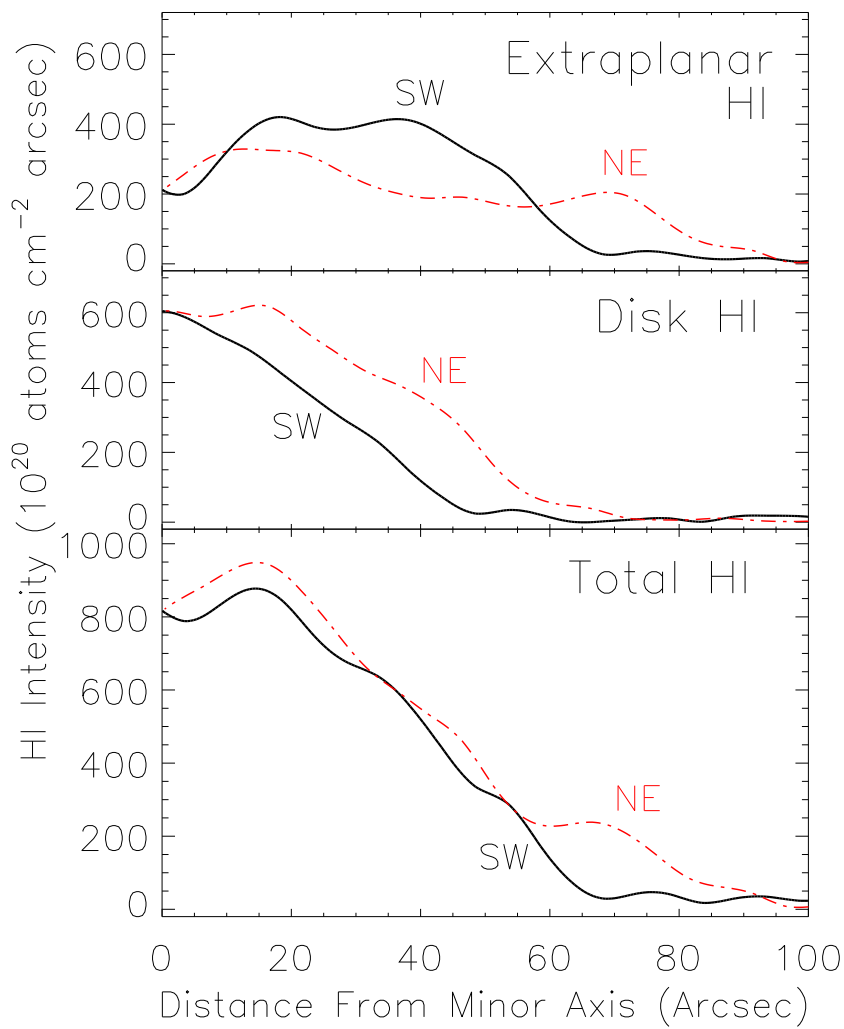}
\caption{
HI intensity slices along and parallel to the major axis of NGC~4522.
a.) Extraplanar  component is the sum of all HI emission above the major axis 
from +11$''$ to +50$''$.
b.) Disk component is the sum of all HI emission along major axis from 
-20$''$ to +11$''$.
c.) Total is the sum of both extraplanar and disk emission.
The SW side, which is rotating into the oncoming ICM, 
has weaker disk emission and stronger extraplanar emission.
However, the sum of extraplanar and disk emission is remarkably similar
on the SW and NE sides.
}
\end{figure}

\begin{figure}
\centering
\includegraphics[angle=-90,scale=0.9]{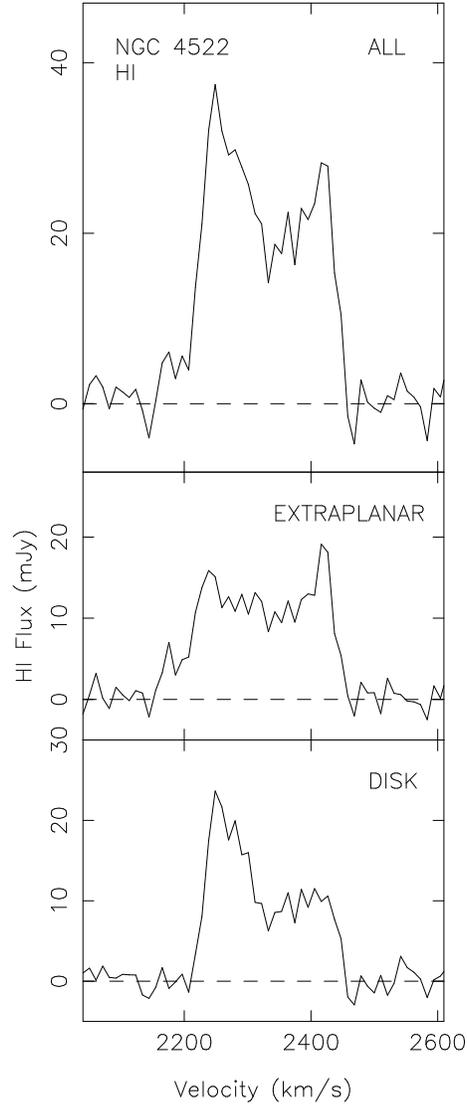}

\caption{
HI spectra for NGC 4522, for whole galaxy, for ``disk'' component
(-20$''$ to +11$''$), and for extraplanar  component (+11$''$ to +50$''$).
The velocity extents of the disk and extraplanar components
are similar, although the extraplanar emission extends to
lower velocities.
The global profile shows a modest velocity asymmetry,
due to a strong asymmetry in the disk,
which has a weak high velocity component.
The extraplanar component is more symmetric,
but with a blueshifted tail.
}
\end{figure}

\begin{figure}
\epsscale{0.95}
\plotone{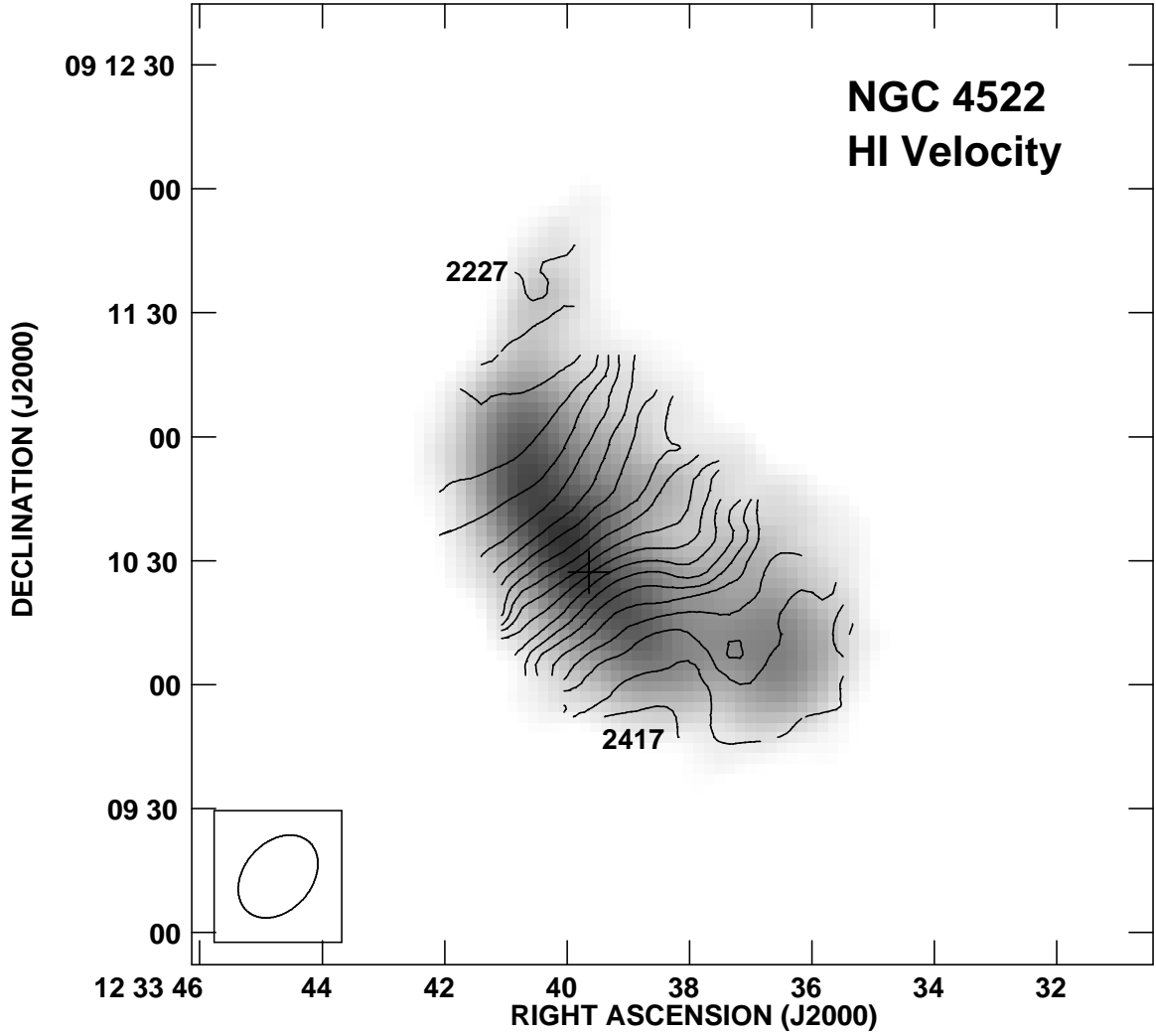}
\caption{
HI velocity field (moment 1) contour map on total intensity (moment 0) 
grayscale map. Cross marks kinematic center, which has velocity of 2337 km s$^{-1}$.
Contour increment 10 km s$^{-1}$.
}
\end{figure}

\begin{figure}
\epsscale{0.95}
\plotone{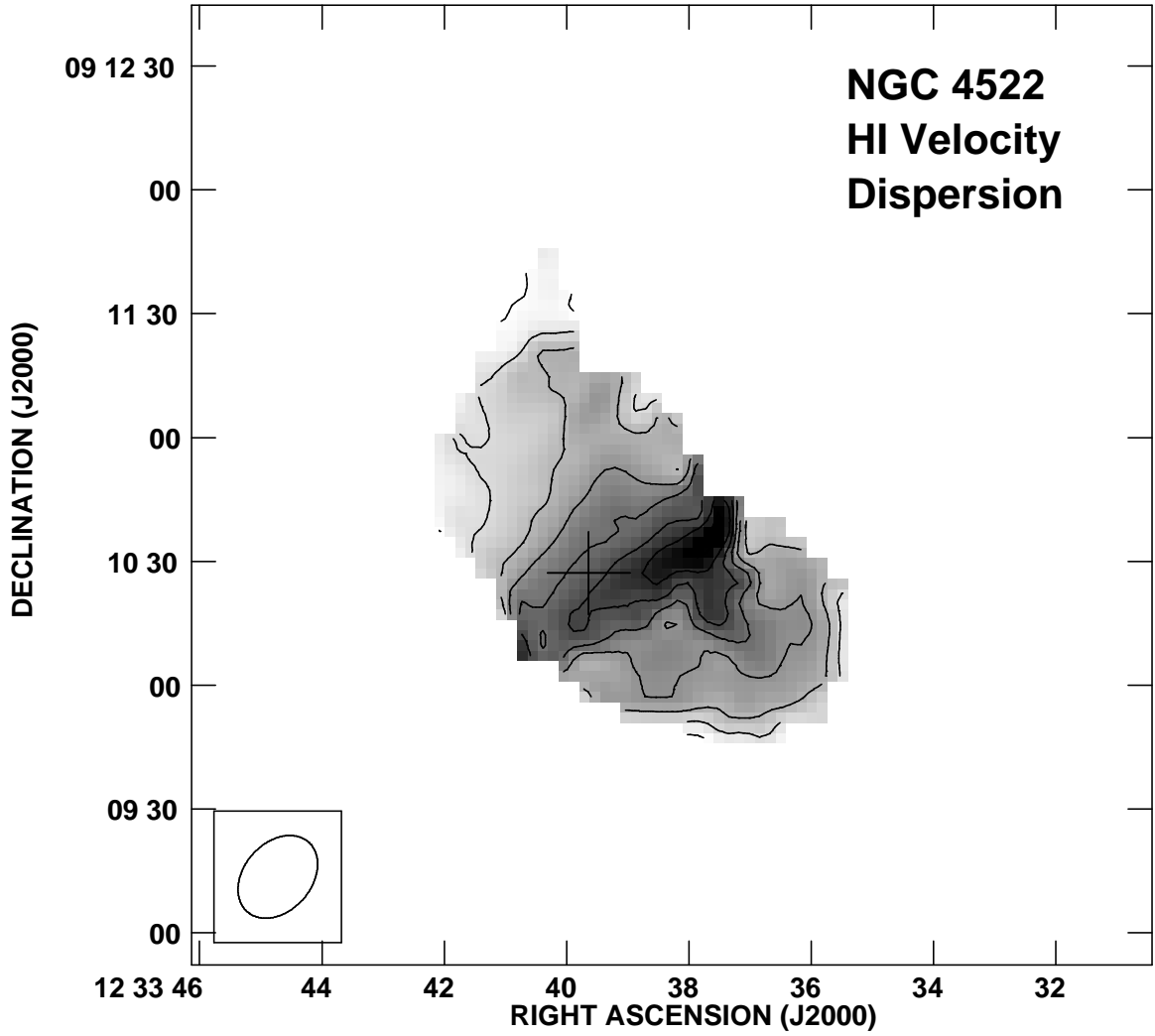}
\caption{
HI velocity width (moment 2) contour and grayscale map.
Contours are 10, 15, 20, 25, 30, 35, 40, and 45 km s$^{-1}$. Cross marks kinematic center.
Note relatively large linewidths to the SW of nucleus in the main disk,
and in the SW extraplanar component.
}
\end{figure}

\begin{figure}
\epsscale{0.85}
\plotone{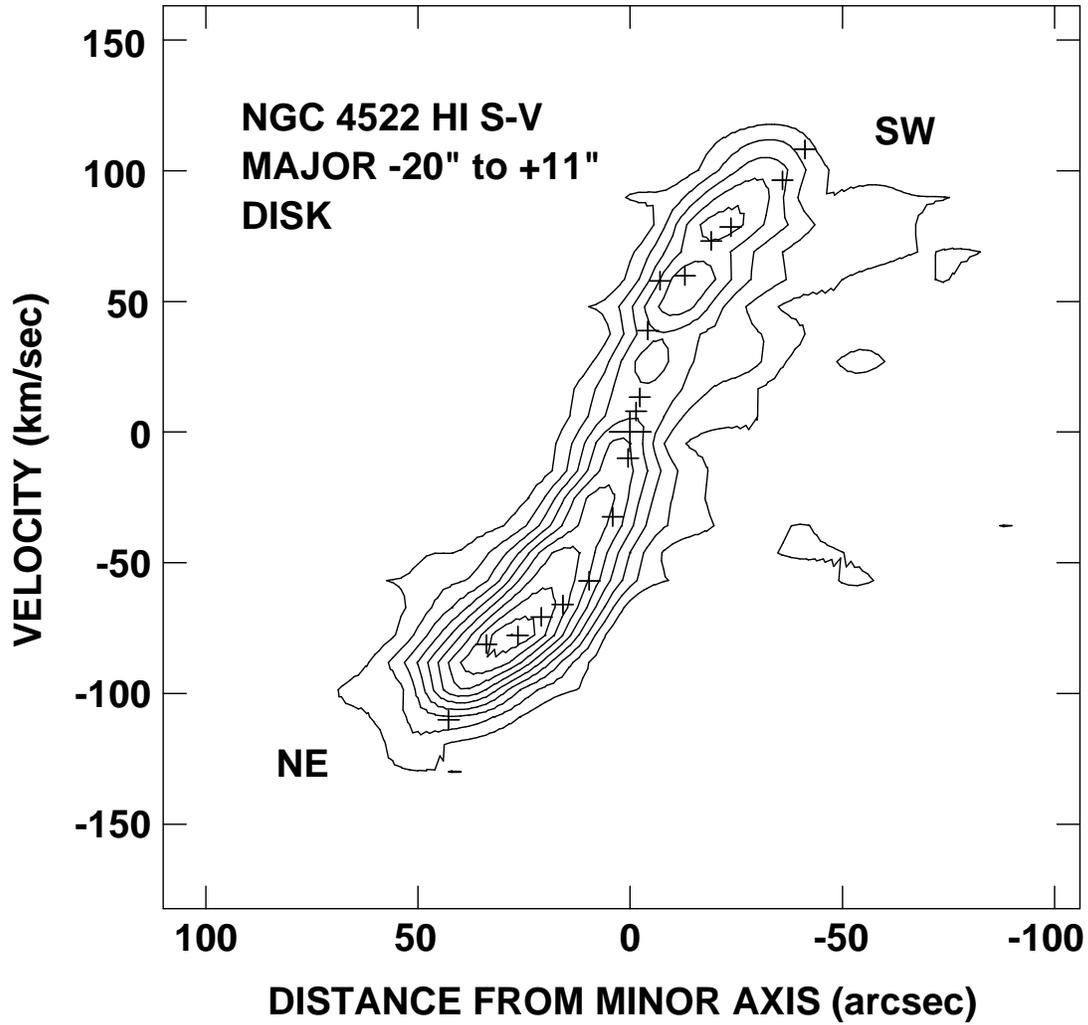}
\caption{
Major axis position-velocity diagram of ``disk'' HI emission, defined as 
between -20$''$ and +11$''$ with respect to the major axis.
Large cross marks kinematic center, and small crosses mark major axis
H$\alpha$ velocities from Rubin \etal (1999).
}
\end{figure}

\begin{figure}
\epsscale{0.95}
\plotone{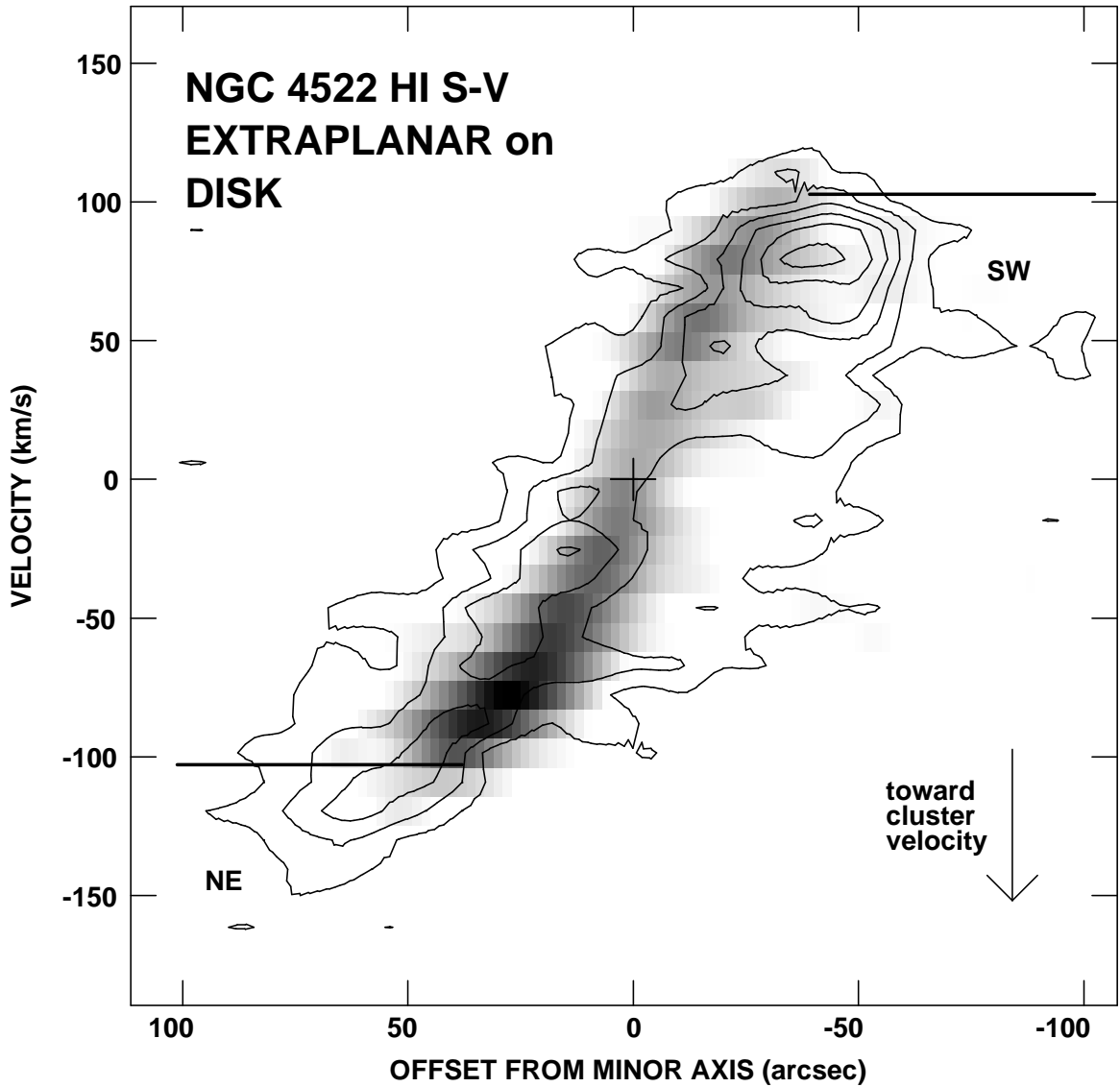}
\caption{
Major axis HI position-velocity diagram for  ``disk'' component, shown as grey scale,
overlain on contour plot of ``extraplanar'' component. ``Disk'' and 
'`extraplanar'' regions are shown in Figure 4.
Solid lines indicate a flat extension of the rotation curve, which peaks at
103 km s$^{-1}$. Note that
much of the extraplanar gas is blueshifted toward the mean cluster velocity.
}
\end{figure}

\begin{figure}
\epsscale{0.95}
\plotone{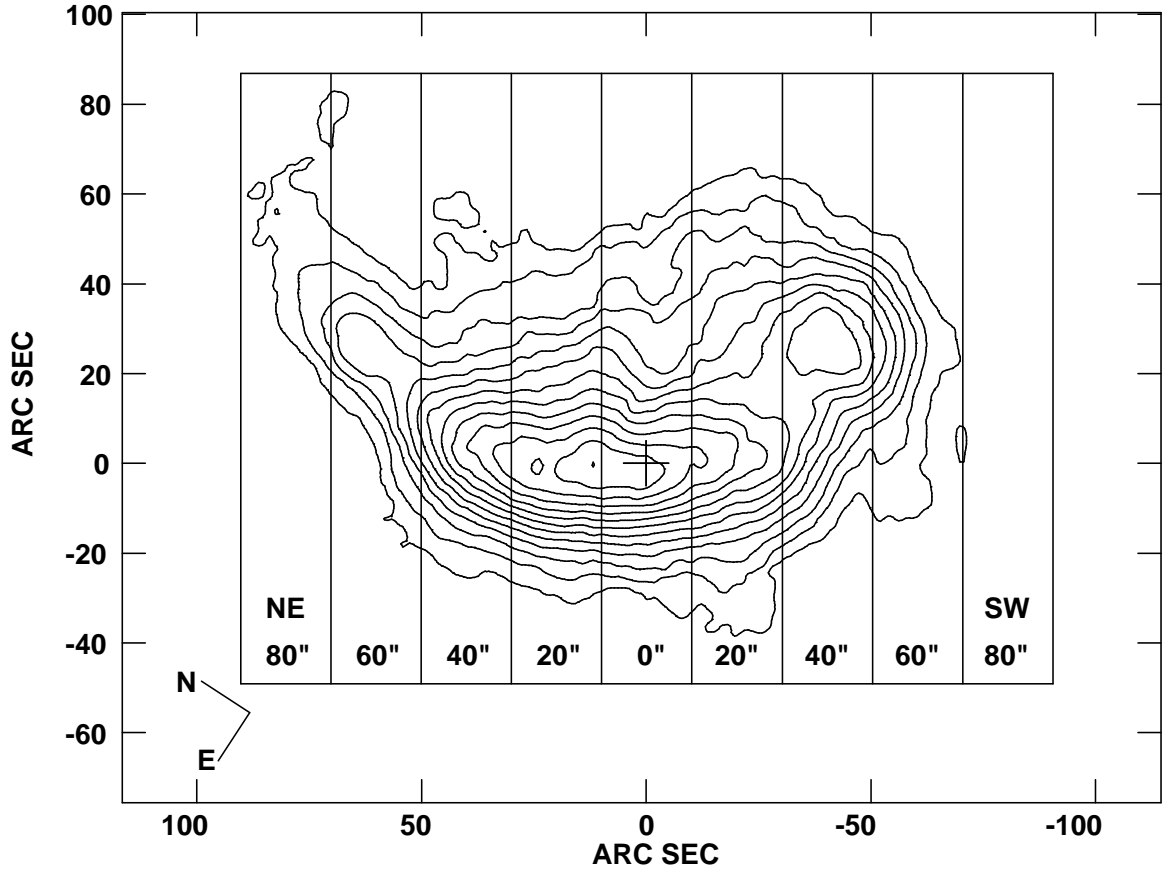}
\caption{HI moment 0 map of NGC~4522, rotated by 57 degrees so that
the major axis is oriented horizontally. The inset boxes 
identify the regions sampled in the associated position-velocity plots
shown in Figure 12,
which are from cuts perpendicular to the major axis of the galaxy.
}
\end{figure}

\begin{figure}
\epsscale{0.28}
\plotone{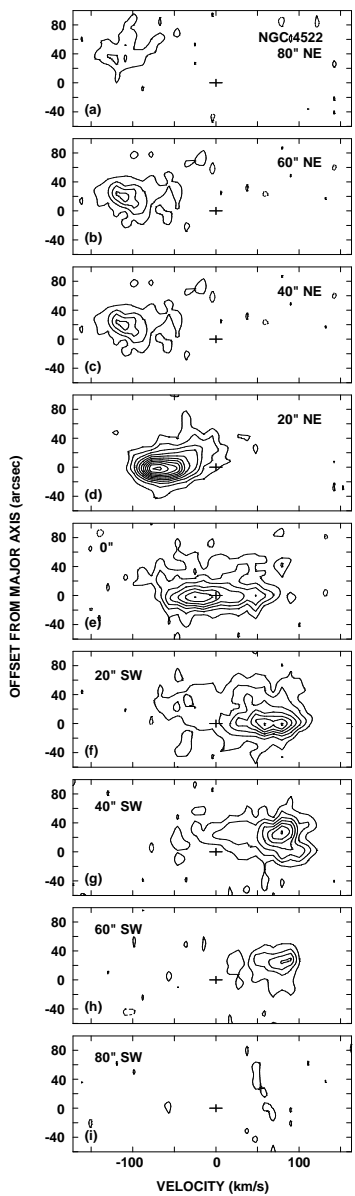}
\caption{
HI position-velocity plots for NGC 4522,
perpendicular to the major axis of the galaxy.
Regions covered in each panel correspond to the 
regions marked in Figure 11.
Each slice is 20$''$ wide and independent
of the other slices.
The middle panel (e) shows a cut along the minor axis of the galaxy, 
and includes emission within $\pm$10$''$ of the minor axis.
The origin corresponds to the HI kinematic center.
Positive spatial offsets correspond to the extraplanar emission.  
Note in particular the velocity structure of
the extraplanar component at 40'' SW in panel g. The line has a total width
of 150 km s$^{-1}$. The strongest component has a peak velocity which
is about 10 km s$^{-1}$ blueshifted from the corresponding disk component,
but fainter emission extends blueward for about 100 km s$^{-1}$.
}
\end{figure}

\begin{figure}
\epsscale{0.94}
\plotone{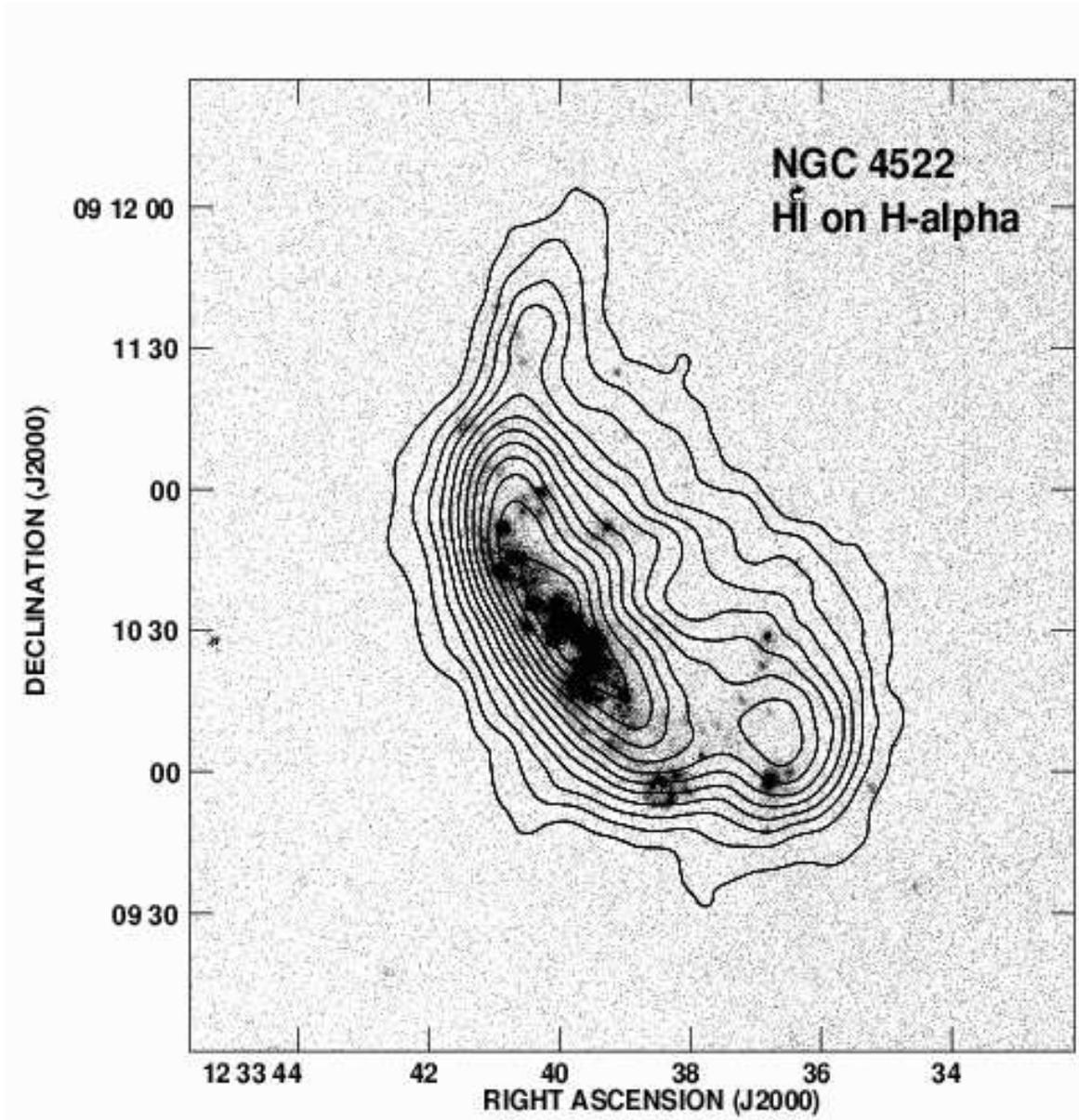}
\caption{HI contour map on H$\alpha$ greyscale image of NGC~4522.
Lowest HI contour level and contour increments are 50 mJy/beam km/s.
The spatial distributions of these 2 ISM components are similar
on scales $\geq$15$''\simeq$1 kpc.
There are HII regions associated with each of the 3 major 
extraplanar HI peaks, and those in the SW are much more luminous.
The outermost bright disk HII complex in the SW appears to form
a 6$''$=500 pc diameter bubble.
}
\end{figure}

\begin{figure}
\epsscale{0.85}
\plotone{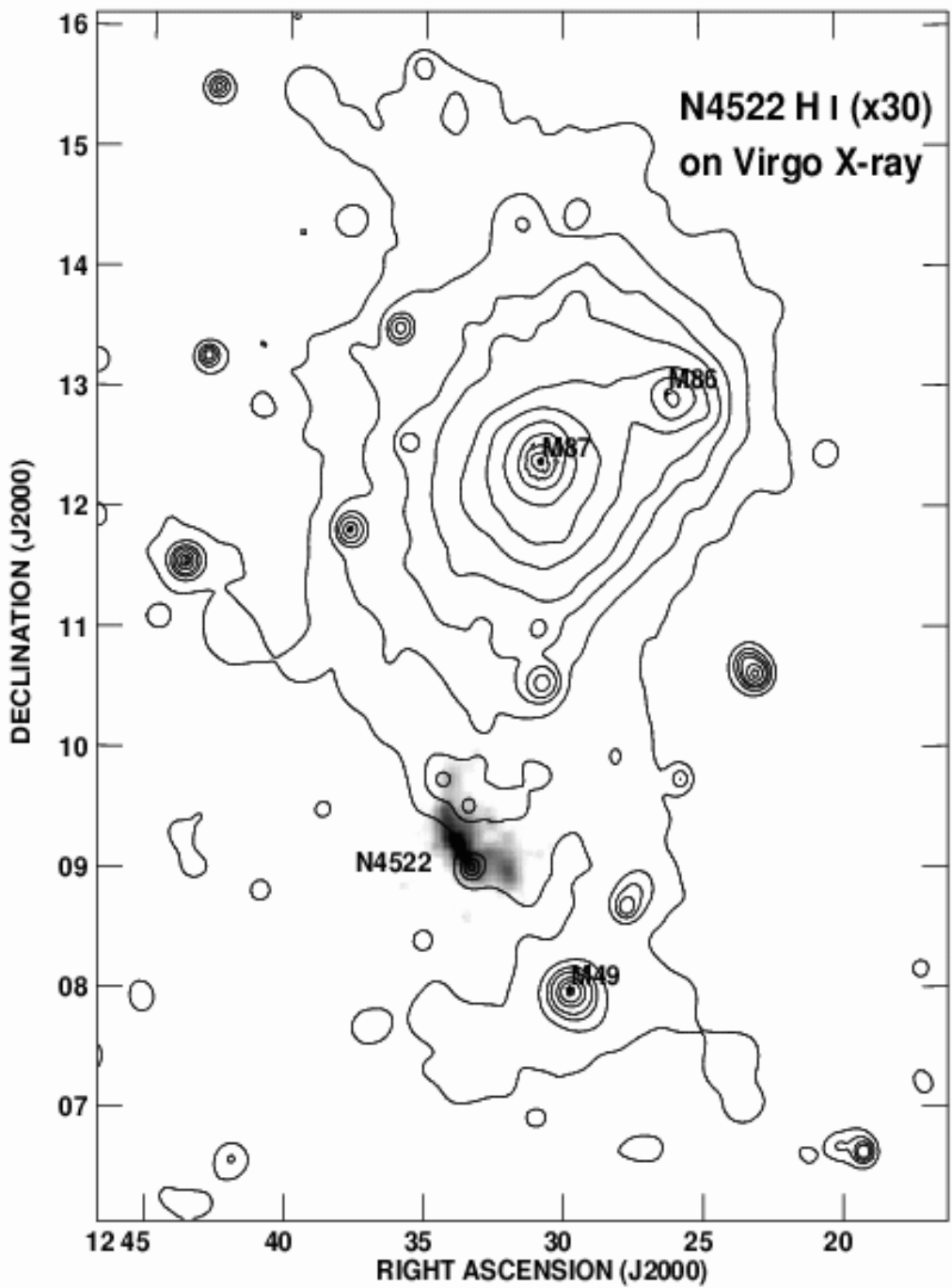}
\caption{HI greyscale map of NGC~4522, scaled up in size by a factor of 30,
on a ROSAT X-ray map of the Virgo cluster from B\"{o}hringer \etal (1994). 
Logarithmic contours intervals are
0,0.1,0.2,0.3,0.5,0.7,1.0,1.3,1.6,2,2.5 counts/s/pixel.
The map indicates the locations of the giant ellipticals M87, M86, and M49, 
which are each associated with sub-clusters.}
\end{figure}

\eject

\end{document}